\documentclass[%
 reprint,
 amsmath,amssymb,
 aps,
superscriptaddress,
showpacs,preprintnumbers,
twocolumn,
tightenlines,
prb,
]{revtex4-2}

\usepackage{graphicx}
\usepackage{dcolumn}
\usepackage{bm}%
\usepackage{dcolumn}
\usepackage{siunitx}
\usepackage{amssymb}
\usepackage{booktabs}
\usepackage{setspace}
\usepackage{xcolor}

\begin{document}

\title{Hydrostatic pressure effects in the Kitaev quantum magnet $\alpha$-RuCl$_3$: A single-crystal neutron diffraction study}

\author{Xiao Wang}
\thanks{These authors contributed equally to this work}
\affiliation{J\"ulich Centre for Neutron Science (JCNS) at Heinz Maier-Leibnitz Zentrum (MLZ), Forschungszentrum J\"ulich, Lichtenbergstrasse 1, D-85747 Garching, Germany}

\author{Fengfeng Zhu}
\thanks{These authors contributed equally to this work}
\affiliation{J\"ulich Centre for Neutron Science (JCNS) at Heinz Maier-Leibnitz Zentrum (MLZ), Forschungszentrum J\"ulich, Lichtenbergstrasse 1, D-85747 Garching, Germany}
\affiliation{State Key Laboratory of Functional Materials for Informatics, Shanghai Institute of Microsystem and Information Technology, Chinese Academy of Sciences, 200050 Shanghai, China}

\author{Navid Qureshi}
\affiliation{Institut Laue-Langevin, 71 Avenue des Martyrs, 38042 Grenoble Cedex 9, France}

\author{Ketty Beauvois}
\affiliation{Institut Laue-Langevin, 71 Avenue des Martyrs, 38042 Grenoble Cedex 9, France}

\author{Junda Song}
\affiliation{J\"ulich Centre for Neutron Science (JCNS) at Heinz Maier-Leibnitz Zentrum (MLZ), Forschungszentrum J\"ulich, Lichtenbergstrasse 1, D-85747 Garching, Germany}

\author{Thomas Mueller}
\affiliation{J\"ulich Centre for Neutron Science (JCNS) at Heinz Maier-Leibnitz Zentrum (MLZ), Forschungszentrum J\"ulich, Lichtenbergstrasse 1, D-85747 Garching, Germany}

\author{Thomas Br\"uckel}
\affiliation{J\"ulich Centre for Neutron Science (JCNS) and Peter Grünberg Institut (PGI), JARA-FIT, Forschungszentrum J\"ulich, D-52425 J\"ulich, Germany} 

\author{Yixi Su}
\email{y.su@fz-juelich.de}
\affiliation{J\"ulich Centre for Neutron Science (JCNS) at Heinz Maier-Leibnitz Zentrum (MLZ), Forschungszentrum J\"ulich, Lichtenbergstrasse 1, D-85747 Garching, Germany}

\date{\today}

\begin{abstract}
We report a comprehensive single-crystal neutron diffraction investigation of the Kitaev quantum magnet $\alpha$-RuCl$_{3}$ under hydrostatic pressure. Utilizing a He-gas pressure cell, we successfully applied an ideal hydrostatic pressure \textit{in situ} at low temperatures, which allows to effectively eliminate any possible influences from the structural transition occurring between 200 K and 50 K under ambient conditions. Our experiments reveal a gradual suppression of the ziagzag antiferromagnetic order as hydrostatic pressure increases. Furthermore, a reversible pressure-induced structural transition occurs at a critical pressure of $P_d$ = 0.15 GPa at 30 K, as evidenced by the absence of magnetic order and non-uniform changes in lattice constants. The decrease in magnetic transition temperature is discussed in relation to a pressure-induced change in the trigonal distortion of the Ru-Cl octahedra in this compound. Our findings emphasize the significance of the trigonal distortion in Kitaev materials, and provide a new perspective on the role of hydrostatic pressures in the realization of the Kitaev quantum spin liquid state in $\alpha$-RuCl$_{3}$.\\

\end{abstract}

\date{\today}

\maketitle

The celebrated Kitaev model has garnered significant research attention over the past decade due to its exact solvability, the presence of the Kitaev quantum spin liquid (KQSL) ground state, and the emergence of novel topological order with exotic fractionalized Majorana-like excitations, which are promising for potential applications in quantum computing \cite{Kitaev2006}. The original Kitaev model is formulated based on a spin-1/2 honeycomb lattice with three types of bond-direction dependent nearest-neighbor interactions. Based on this concept, Jackeli and Khaliullin proposed that such a model could be realized in 4$d$ or 5$d$ transition metal compounds, such as Na$_2$IrO$_3$, $\alpha$-Li$_2$IrO$_3$, and $\alpha$-RuCl$_3$ \cite{Jackeli2009,Plumb2014,Yamaji2014,Yadav2016,Banerjee2017,Do2017,Kim2020}. In these compounds, the magnetic ions Ir$^{4+}$ (5$d^5$) or Ru$^{3+}$ (4$d^5$) reside within edge-sharing octahedra formed by the O$^{2-}$ or Cl$^{-}$ ligands. The crystal field effect causes the splitting of $d$ orbitals into a doubly degenerate $e_{g}$ manifold and a triply degenerate $t_{2g}$ manifold with an effective angular momentum of $\mathbf{L}_{\text {eff }} = 1$ \cite{Rau2016,Hermanns2018,Takagi2019,Khomskii2021}. The $t_{2g}$ manifold is further split by the strong spin-orbit coupling (SOC) $\lambda\mathbf{L} _{\text {eff }}\cdot \mathbf{S}$, resulting in a $j_{\text {eff }}=1 / 2$ doublet ground state. The ligand-mediated nearest-neighbor exchange interaction of the $j_{\text {eff }}=1 / 2$ magnetic ions with three nearest magnetic ions is generally highly anisotropic due to the spin-orbit entanglement and can be described by the bond-direction dependent Kitaev model \cite{Jackeli2009,Plumb2014,Yamaji2014,Yadav2016,Banerjee2017,Do2017,Kim2020,Singh2012}.

However, mainly due to the presence of significant non-Kitaev interactions, all of the above-mentioned candidate materials exhibit conventional long-range magnetic order at low temperatures \cite{Chaloupka2010,Sizyuk2014,Rau2014,Rau2014_2,Sears2015,Chaloupka2016,Sizyuk2016,Rousochatzakis2018}. It is thus necessary to include those non-Kitaev interactions, such as the isotropic Heisenberg interaction $J_{1}$ and the off-diagonal interaction $\Gamma$, which originate from the direct overlap among the spatially extended distribution of $d$ orbitals, in the so-called extended $K$-$J_{1}$-$\Gamma$ model \cite{Rau2014}. If the surrounding cubic ligand-coordination environment of magnetic ions is distorted, additional contributions due to long-range or further nearest-neighbor exchange interactions, namely $\Gamma^{\prime}$ and $J_{3}$, potentially need to be included \cite{Sizyuk2014,Rau2014_2,Tovar2022}. For instance, the occurrence of a zigzag antiferromagnetic order below approximately 7.5 K under zero magnetic field in $\alpha$-RuCl$_3$ can be explained by the minimal $K$-$J_1$-$\Gamma$-$\Gamma^{\prime}$ or $K$-$J_1$-$\Gamma$-$J_{3}$ model \cite{Winter2016,Winter2017,Sears2020,Suzuki2021}. Intriguingly, the zigzag antiferromagnetic order can be completely suppressed by a magnetic field of approximately 8 T applied along the hexagonal $a$ axis \cite{Sears2017,Baek2017}, leading to a possible field-induced KQSL state that is signalized by the emergence of a half-integer quantized thermal Hall effect \cite{Kasahara2018}. While the exact nature of the field-induced quantum states in $\alpha$-RuCl$_3$ is still subject to an ongoing and extensive research effort in condensed matter physics community \cite{Kasahara2018,Czajka2021,Bruin2022,Tanaka2022,Czajka2023}, due to the apparent fragility of the zigzag magnetic order and a proximate KQSL behavior in this compound, it has been clearly shown that magnetic fields can be employed for the efficient tuning of the magnetic states in the Kitaev materials. 

In addition to magnetic fields, both uniaxial and hydrostatic pressures are also widely used as an efficient tuning parameter for the studies of quantum phase transitions as well as possible pressure effects in quantum magnets and unconventional superconductors \cite{Rueegg2008,Gati2020}. The applied pressure can significantly impact the structural, electronic and magnetic properties of these materials, often driving them towards different quantum phases, including potential quantum spin liquid states. The generation of high pressure could usually be achieved with either a purposely made uniaxial press, or clamp pressure cell or diamond anvil cell using suitable pressure transmitting fluids. A number of experimental investigations, including magnetic susceptibility, thermal expansion, NMR, infrared and Raman spectroscopy and X-ray diffraction, of hydrostatic pressure effects on both magnetism and structure of $\alpha$-RuCl$_{3}$ have been carried out recently \cite{Cui2017,He2018,Wang2018,Bastien2018,Biesner2018,Li2019,Wolf2022}. However, conflicting results have been reported by different groups, particularly concerning the role of hydrostatic pressures on the zigzag antiferromagnetic order and on possible pressure-induced structural changes. This is largely owing to the challenge to achieve an ideal hydrostatic pressure condition, and meanwhile, to allow the access from a suitable experimental probe in a pressure dependent investigation.

In this letter, we report the first comprehensive single-crystal neutron diffraction study of hydrostatic pressure effects in $\alpha$-RuCl$_3$ using a He-gas pressure cell. This experimental setup not only realizes an ideal hydrostatic pressure condition, but also allows \textit{in situ} pressure-changing at low temperatures, thus paving the way for a detailed microscopic study of the pressure-temperature phase diagram of both magnetism and structure in $\alpha$-RuCl$_3$ via single-crystal neutron diffraction. Our results compellingly demonstrate that the zigzag magnetic order transition temperature can be reduced with the application of hydrostatic pressures, and a pressure-induced reversible structural transition occurs before the complete suppression of the magnetic order. Our study also reveals a compression effect on the $c$ axis, which modifies the trigonal distortion and drives the system towards the Kitaev exchange interaction dominant side.

High-quality single-crystal samples of $\alpha$-RuCl$_3$ were synthesized using the chemical vapor transport (CVT) method \cite{He2018,Mi2021}. The single-crystal neutron diffraction experiment was carried out at the four-circle diffractometer D10 at the Institut Laue-Langevin (ILL), Grenoble, with an incident wavelength of 2.36 \AA. A pyrolytic graphite (PG) filter is used for the suppression of higher-order harmonics. Two detector configurations were used in the experiment, either with a two-dimensional (2D) microstrip detector for an efficient measurement over a large area in reciprocal space, or with a single $^3$He detector that is combined with the triple-axis mode for high-resolution and low-background measurements. A truly hydrostatic pressure up to 7 kbar can be generated by a He-gas compressor connected to a dedicated TiZr pressure cell. A pressure gauge is used to monitor the generated pressure directly. A 60 mg high-quality single crystal of $\alpha$-RuCl$_3$ with dimensions of $\sim$0.5$\times$5$\times$7 mm$^3$ was aligned in the ($h$,0,$l$) scattering plane and attached to a thin aluminum plate for measurement. A small MgO single crystal that was mounted adjacent to the sample is used for the further \textit{in situ} calibration of hydrostatic pressure. The He-gas pressure cell is attached to a dedicated sample stick and is then inserted into a standard $^{4}$He ILL Orange Cryostat. The sample was first cooled down to 1.5 K, and then the temperature-dependent data at zero pressure were collected by gradually heating the sample up to 10 K. For the measurements of the magnetic reflections at each pressure, the sample was first heated up to 30 K, and the hydrostatic pressure was carefully applied at this temperature, followed by cooling the sample back down to 1.5 K and then carrying out the measurements at that pressure. Applying hydrostatic pressures at 30 K effectively eliminates any possible influences of the pronounced structural transition in the intermediate temperature range (i.e. roughly between 200 K and 50 K) in $\alpha$-RuCl$_3$ \cite{Sears2015,Park2016,Johnson2015,Cao2016}, which has been found in many CVT-grown single crystals and also in our sample, but its pressure-dependent behavior remains unclear \cite{Kubota2015,He2018,Zhou2019,Gass2022,Reschke2018}. Since all the observed reflections in our neutron diffraction experiment can be well indexed based on a rhombohedral crystal structure model \cite{Park2016}, we have employed the $R\overline{3}$ space group for the low-temperature structure of $\alpha$-RuCl$_3$ in this work.

\begin{figure}[htbp]
	\centering
	\includegraphics[width=8cm]{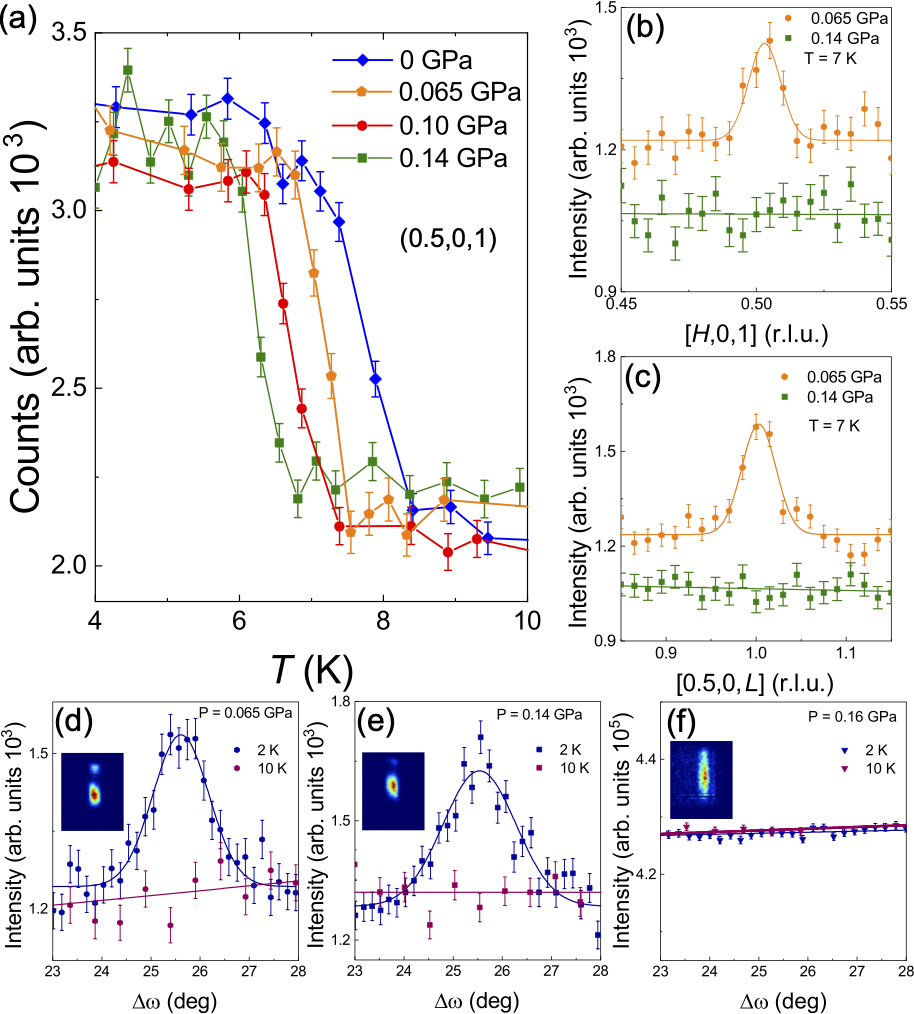}
	\caption{\label{fig:Fig_1}
	Hydrostatic pressure effect on the zigzag antiferromagnetic order in $\alpha$-RuCl$_3$. (a) Temperature dependence of the peak intensity of the magnetic reflection (0.5,0,1) under different hydrostatic pressures.
	(b) and (c) $Q$-scans ($H-$ and $L$-scans) of the magnetic reflection (0.5,0,1) under hydrostatic pressures of 0.065 GPa and 0.14 GPa at 7 K, respectively. Each data point under 0.065 GPa and 0.14 GPa was measured for 6 minutes, and under 0.16 GPa was measured for 20 minutes to ensure good data quality.
	(d-f) Rocking curves for the magnetic reflection (0.5,0,1) under hydrostatic pressures of 0.065 GPa, 0.14 GPa, and 0.16 GPa at both 2 K and 10 K, respectively. The inset shows the corresponding intensity mapping of the nuclear Bragg reflection (-3,0,0) obtained with the 2D area detector. Note that the weak feature above the main peak in the 2D intensity mapping in the inset is due to a minority crystalline grain in the sample, which does not affect the interpretation and results reported in this work. 
	}
\end{figure}

The results from our neutron diffraction study of the zigzag antiferromagnetic order under various temperatures and pressures are shown in Fig.\ref{fig:Fig_1}. The temperature dependence of the magnetic reflection (0.5,0,1) is shown in Fig.\ref{fig:Fig_1}(a). Under zero pressure, the zigzag antiferromagnetic order transition occurs at 7.91(5) K. With increasing hydrostatic pressure, it is evident that the zigzag order transition temperature decreases systematically to 6.30(3) K under 0.14 GPa. This can be further verified by comparing the $Q$-scans of the magnetic reflection (0.5,0,1) under 0.065 GPa and 0.14 GPa at 7 K in Fig.\ref{fig:Fig_1}(b) and (c), respectively. (0.5,0,1) can be clearly observed under 0.065 GPa, while it is completely absent under 0.14 GPa at 7 K. This result is quite intriguing, as following the large linear slope of -11.5 K/GPa from 0 GPa to 0.14 GPa, it hints that the zigzag order could be fully suppressed by an extraordinarily small pressure of approximately 0.7 GPa, potentially driving $\alpha$-RuCl$_3$ into the KQSL state. However, this possibility is not realized in this experiment due to a pressure-induced structural transition at higher pressures. As illustrated in Fig.\ref{fig:Fig_1}(d-f), the rocking curves of the magnetic reflection (0.5,0,1) and the corresponding nuclear Bragg reflection (-3,0,0), measured at 2 K and 10 K under both 0.065 GPa and 0.14 GPa, respectively, exhibit a well-defined peak at 2 K, confirming that the crystal structure is retained along with the zigzag order up to 0.14 GPa. However, at 0.16 GPa, the magnetic reflection (0.5,0,1) vanishes along with the significantly broadened (-3,0,0), indicating the occurrence of a pressure-induced structural transition in the sample.

\begin{figure}[htbp]
	\centering
	\includegraphics[width=8cm]{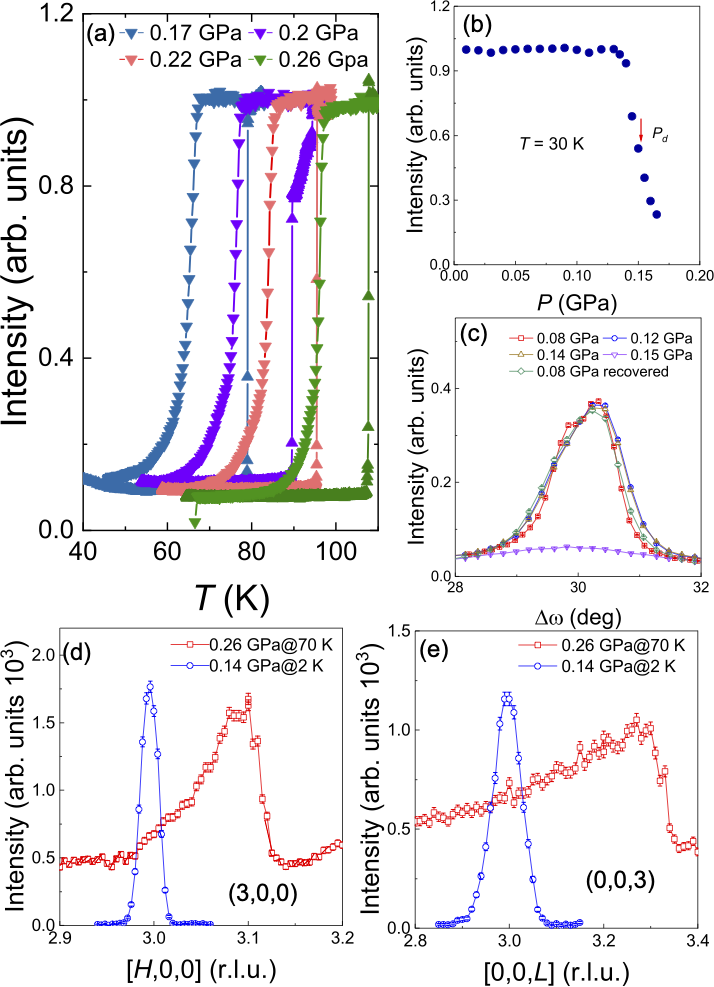}
	\caption{\label{fig:Fig_2}
		Pressure-induced structural transition. (a) Temperature dependence of the normalized integrated intensity of the structural reflection (-3,0,0) under different hydrostatic pressures upon cooling ($\blacktriangledown$) and heating ($\blacktriangle$). The result was obtained via the integration over the intensity mapping from the 2D area detector, and the subsequent normalization by the intensity of the same refection in the undistorted structure under zero pressure.
		(b) Hydrostatic pressure dependence of the normalized integrated intensity of the structural reflection (-3,0,0) at 30 K. The data were obtained by carefully increasing the He-gas pressure \textit{in situ} at 30 K, and by the same intensity integration and normalization procedure as in (a).
		(c) Rocking curves of the structural reflection (-3,0,0) under different pressures.
		(d-e) $Q$-scans along the $H$ and $L$ reciprocal directions for the structural reflections (3,0,0) and (0,0,3) under both 0.14 GPa and 0.26 GPa at 70 K and 2 K, respectively. The data shown in (c-e) were obtained with the single $^3$He detector combined with the triple-axis mode.
	}
\end{figure}

The neutron diffraction investigations of the pressure-induced structural transition are summarized in Fig.\ref{fig:Fig_2}. Our measurement protocol is that each hydrostatic pressure is loaded at 100 K, and after pressure loading, the sample is cooled down and then warmed up slowly while the intensity mapping of the structural Bragg reflection (-3,0,0) is recorded with the 2D area detector simultaneously. The temperature dependence of the normalized integrated intensity of (-3,0,0) under different pressures is shown in Fig.\ref{fig:Fig_2}(a). It is obvious that the pressure-induced structural transition temperature increases gradually with the increase of the hydrostatic pressure, and a significant hysteresis of approximately 15 K exists between the heating and cooling of the sample, suggesting the first-order nature of this pressure-induced structural transition. To clarify the critical pressure at 30 K, the integrated intensity of (-3,0,0) was measured while increasing the pressure gradually, and its normalized result is shown in Fig.\ref{fig:Fig_2}(b). From around 0.14 GPa, the intensity of (-3,0,0) drops dramatically, we could thus define the critical pressure $P_{d}$ = 0.15 GPa where the intensity is decreased to its half maximum. Fig.\ref{fig:Fig_2}(c) shows the rocking curves of the (-3,0,0) reflection at different pressures. It can be noticed that under 0.15 GPa, the (-3,0,0) reflection almost disappears in the rocking-curve scan, but it still retains half of its maximum intensity in the 2D integrated intensity as shown in (b). This implies a significant broadening of the (-3,0,0) reflection under this pressure. The $Q$-scans of the (3,0,0) reflection under hydrostatic pressures of 0.14 GPa and 0.26 GPa are plotted in Fig.\ref{fig:Fig_2}(d). Under 0.26 GPa, the peak shifts to a higher-$Q$ position and becomes much broader compared to that under 0.14 GPa, suggesting a sudden shrinkage in the hexagonal in-plane lattice constant. This behavior is also observed for the $Q$-scans of the (0,0,3) reflection, which indicates a non-uniform compression of the $c$ axis under 0.26 GPa, as shown in Fig.\ref{fig:Fig_2}(e). Very interestingly, we have also found that this pressure-induced structural transition is reversible, i.e., after releasing the pressure, the ambient-pressure crystal structure is fully recovered, which can be clearly demonstrated by the same rocking curves of the (-3,0,0) reflection under 0.08 GPa measured for the two different cases respectively in loading/off-loading of hydrostatic pressures, as shown in Fig.\ref{fig:Fig_2}(c).

\begin{table}[htbp]
	\footnotesize
	\caption[structure refinement results]{\label{tab:comp}
		Comparison of the determined crystal structures of $\alpha$-RuCl$_3$ at 2 K under 0 GPa and 0.14 GPa, respectively.
	}
	\begin{center}
		\begin{tabular}{p{3.3cm}p{1.3cm}p{1.3cm}p{1.8cm}}
			\hline
			\hline
			&0 GPa &0.14 GPa&Change ratio \\
			\hline
			Ru-Ru($c$) (\AA)&5.648(5)&5.633(4)&-0.266\% \\
			Ru-Ru($ab$ nearest) (\AA)&3.446(3)&3.443(3)&-0.087\%\\
			$\angle$Ru-Cl-Ru (deg) &93(3)&94(2)&1.08\% \\
			\hline
			\hline
		\end{tabular}
	\end{center}
\end{table}

\begin{figure}[]
	\centering
	\includegraphics[width=8cm]{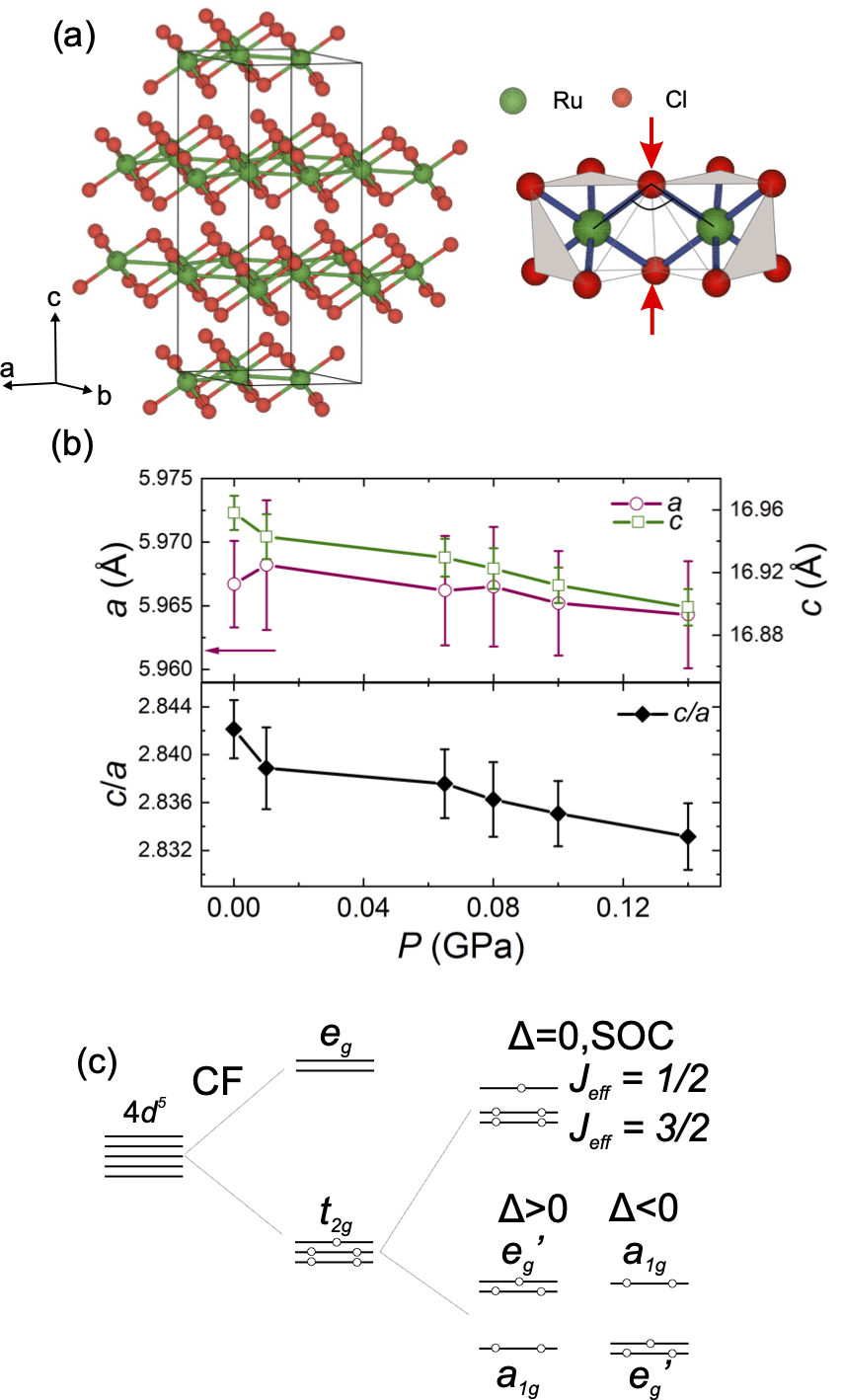}
	\caption{\label{fig:Fig_3}
		(a) Left: The crystal structure model (R$\overline{3}$) of $\alpha$-RuCl$_3$ at low temperatures and under the ambient pressure condition. Right: The illustration of the Cl-Ru-Cl bond-angle and the compression of the $c$ axis.
		(b) Lattice constants $a$, $c$ of $\alpha$-RuCl$_{3}$ and the corresponding $c$/$a$ ratio as a function of hydrostatic pressure.
		(c) Effects of crystal field splitting (CF), spin-orbit coupling (SOC) and trigonal distortion ($\Delta$) on the 4$d^{5}$-orbitals states.
	}
\end{figure}

Earlier studies of $\alpha$-RuCl$_{3}$ under hydrostatic pressures have revealed a collapsed phase with a structural dimerization of the nearest Ru-Ru bonds above a critical pressure \cite{Biesner2018,Bastien2018,Li2019}, which is consistent with the $Q$-scan of (3,0,0) reflection, where an abrupt contraction of the lattice constant $a$ is observed. Such a similar pressure-induced collapsed phase was also found in other Kitaev materials, e.g., $\alpha$-Li$_{2}$IrO$_{3}$ \cite{Shen2022} and $\beta$-Li$_{2}$IrO$_{3}$ \cite{Takayama2019}, in which the nearest Ir-Ir bonds are structurally dimerized under pressures. Thus, we could assign the name "\textit{structural dimerization}" to the high-pressure phase. Due to the experimental limitation, we could only have access to the reflections in the ($h$,0,$l$) plane; therefore, it is impractical to comprehensively determine the pressure-induced structures in this experiment. Nonetheless, we have manged to collect 62 and 82 nuclear Bragg reflections at 0 GPa and 0.14 GPa, respectively. While there still exists a controversy about the low-temperature crystal structure of $\alpha$-RuCl$_3$ \cite{Park2016,Johnson2015,Cao2016}, our best refinement of the crystal structure via Jana2006 is based on the space group $R\overline{3}$ \cite{Park2016}. Some essential structural parameters, such as the lattice constant $c$, the nearest Ru-Ru bond distance, and the Ru-Cl-Ru bond-angle, are listed in Tab.\ref{tab:comp}. At first glance, the change of the lattice constants along the $c$ direction is rather suspicious, which is about three times larger than the change in the hexagonal lattice constant $a$. This difference can be traced back to the quasi-2D nature of the crystal structure of $\alpha$-RuCl$_{3}$, as illustrated in Fig.\ref{fig:Fig_3}(a). It is well established that $\alpha$-RuCl$_{3}$ is a layered compound, and each layer is weakly bonded by van der Waals interactions, which are much weaker compared to the covalent bonds between Ru$^{3+}$ and Cl$^{-}$ ions. Consequently, the Young's modulus of the $c$ axis is much smaller than that of the $a$ axis, causing a noticeable change of the compressive strain in the $c$ direction under the same pressure. The hydrostatic pressure dependence of the lattice constants $a$ and $c$, as well as the corresponding $c$/$a$ ratio, indicate thus a decreasing tendency with increasing pressures, as plotted in Fig.\ref{fig:Fig_3}(b).

\begin{figure}[]
	\centering
	\includegraphics[width=8cm]{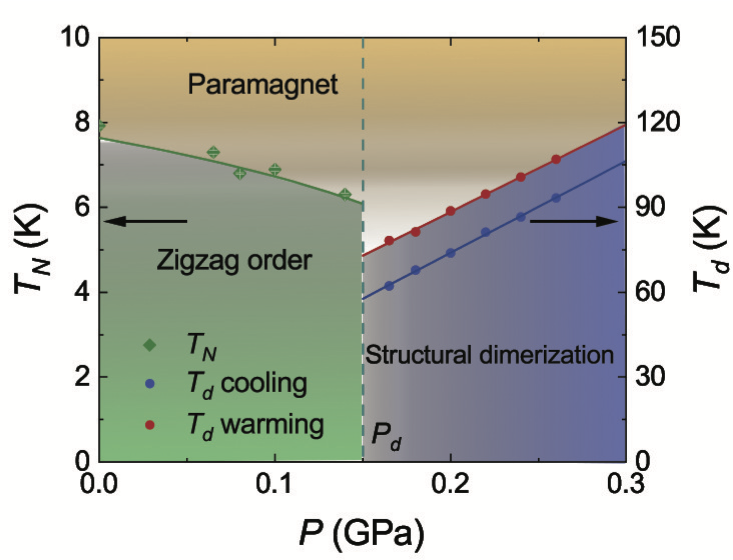}
	\caption{\label{fig:Fig_4}
		Phase diagram of $\alpha$-RuCl$_{3}$ under hydrostatic pressures, which is obtained based on our single-crystal neutron diffraction measurement using a He-gas pressure cell.
		The left $y$-axis refers to the magnetic order transition temperature, and the right $y$-axis refers to the pressure-induced structural transition temperature.
	}
\end{figure}

By combining the pressure dependence of both the zigzag antiferromagnetic order and the pressure-induced collapsed structural phase obtained from our single-crystal neutron diffraction study, a pressure-temperature phase diagram of $\alpha$-RuCl$_{3}$ is constructed, as shown in Fig.\ref{fig:Fig_4}. Given that the hydrostatic pressures were applied at 30 K, our experiment reveals the genuine outcome of hydrostatic pressures on $\alpha$-RuCl$_{3}$ for the first time by avoiding any possible influences of the structural transition in the ambient pressure condition at high temperatures. Besides, in this experiment, a monotonous decrease of the zigzag magnetic order transition by increasing the hydrostatic pressures is unambiguously revealed, which distinguishes itself from the previous macroscopic measurements. Yet interrupted by the pressure-induced structural transition at a small critical pressure of $P_{d}$ = 0.15 GPa, the large slope of -11.5 K/GPa implies a considerable impact of hydrostatic pressures on the zigzag order. When the hydrostatic pressure exceeds $P_{d}$, the $R\overline{3}$ crystal structure is destroyed, and the system enters a reversible dimerization phase, whose transition temperatures experience a linear upward relation with the increasing pressures.

The decrease of the zigzag magnetic order transition temperature as a function of hydrostatic pressures could be interpreted within the framework of the detailed spin Hamiltonian of $\alpha$-RuCl$_{3}$. As discussed, due to the difference in the Young's moduli between the $c$ axis and the hexagonal $ab$ plane, the hydrostatic pressures literally engender a compression along the $c$ axis, which would reshape the trigonal distortion in $\alpha$-RuCl$_{3}$. The existence of the trigonal distortion mainly alters the spin-orbit entangled $j_{\text {eff }}=1 / 2$ ground states of Ru$^{3+}$ ions, but the direction of the distortion, elongation or compression, amends the ground state wave-functions differently, as explained in Fig.\ref{fig:Fig_3}(c). This is particularly true for the $\Gamma^{\prime}$ term, which is intimately connected to the trigonal distortion: an elongation of the octahedra along the $c$ axis ($\Delta$\textgreater 0) causing a positive $\Gamma^{\prime}$, while for the compression ($\Delta$\textless 0), $\Gamma^{\prime}$ is negative \cite{Rau2014_2,Winter2017,Khomskii2021,Liu2022}. Both theoretical and experimental studies have demonstrated that $\alpha$-RuCl$_{3}$ exhibits an elongated octahedral coordination of Ru$^{3+}$ ions, resulting in a positive $\Gamma^{\prime}$. This feature is responsible for the primary differences between $\alpha$-RuCl$_{3}$ and Na$_{2}$IrO$_{3}$ in terms of magnetization and spin excitation properties \cite{Suzuki2021,Liu2022}. In our experiment, the compression of RuO$_6$ octahedra reduces the trigonal distortion, leading to an enhancement of the Kitaev exchange interaction $K$, while decreasing the values of $\Gamma$, $J$, and $\Gamma^{\prime}$ \cite{Liu2022}. This would destabilize the zigzag antiferromagnetic order as the system is driven towards the Kitaev interaction $K$ dominant side.

Moreover, the single-crystal structure refinement of $\alpha$-RuCl$_{3}$ under 0.08 GPa and 0.14 GPa reveals a 1.08\% increase of the Ru-Cl-Ru angle from 93(3)$^\circ$ to 94(2)$^\circ$ in Tab.\ref{tab:comp}. Despite the large uncertainty, this increase is consistent with the physical picture of a substantial compression along the $c$ axis. Theoretical calculations have shown that the Kitaev exchange interaction reaches its maximum at about 94$^\circ$, thus providing another explanation for the decrease of the zigzag magnetic order transition temperature in relation to hydrostatic pressures \cite{Yadav2016}. Furthermore, due to the different origins of exchange interaction parameters in $\alpha$-RuCl$_{3}$, a structurally perfect condition might not be a priority in the realization of KQSL. Achieving KQSL requires a delicate balance among electronic correlations, spin-orbit entanglement, and lattice degree of freedom. Our findings thus emphasize the significance of the trigonal distortion in Kitaev materials and suggest that a potentially straightforward approach to realizing KQSL in $\alpha$-RuCl$_{3}$ might involve applying uniaxial pressures at low temperatures.

In summary, our single-crystal neutron diffraction study of the zigzag magnetic order and the crystal structure of $\alpha$-RuCl$_{3}$ under \textit{in situ} hydrostatic pressures reveals a substantial compression effect on the $c$ axis. This pressure effect modifies the trigonal distortion and drives the system towards the Kitaev exchange interaction dominant side, as indicated by a clear negative correlation between the magnetic transition temperature and hydrostatic pressures. Above $P_{d}$, the low-temperature crystal structure collapses, and the system enters a structural dimerization phase. Our findings highlight the complexity of Kitaev physics and provide a new perspective on the role of hydrostatic pressures in the realization of KQSL in $\alpha$-RuCl$_{3}$.\\

\begin{acknowledgments}

This work is based on the neutron diffraction experiment performed at D10 (ILL, Grenoble). We would like to acknowledge the ILL Sample Environment Group for the help on the operation of He-gas pressure cell at D10. X.W. acknowledges the financial support from the China Scholarship Council (CSC). F.Z. and J.S. acknowledge the financial support from the HGF–OCPC Postdoctoral Program. We would like to acknowledge Susanne Mayr for the help with X-ray Laue.

\end{acknowledgments}


\bibliographystyle{apsrev4-2.bst}
\bibliography{bibtex_rucl3.bib}

\providecommand{\noopsort}[1]{}\providecommand{\singleletter}[1]{#1}%
\begin{thebibliography}{53}%
\makeatletter
\providecommand \@ifxundefined [1]{%
 \@ifx{#1\undefined}
}%
\providecommand \@ifnum [1]{%
 \ifnum #1\expandafter \@firstoftwo
 \else \expandafter \@secondoftwo
 \fi
}%
\providecommand \@ifx [1]{%
 \ifx #1\expandafter \@firstoftwo
 \else \expandafter \@secondoftwo
 \fi
}%
\providecommand \natexlab [1]{#1}%
\providecommand \enquote  [1]{``#1''}%
\providecommand \bibnamefont  [1]{#1}%
\providecommand \bibfnamefont [1]{#1}%
\providecommand \citenamefont [1]{#1}%
\providecommand \href@noop [0]{\@secondoftwo}%
\providecommand \href [0]{\begingroup \@sanitize@url \@href}%
\providecommand \@href[1]{\@@startlink{#1}\@@href}%
\providecommand \@@href[1]{\endgroup#1\@@endlink}%
\providecommand \@sanitize@url [0]{\catcode `\\12\catcode `\$12\catcode
  `\&12\catcode `\#12\catcode `\^12\catcode `\_12\catcode `\%12\relax}%
\providecommand \@@startlink[1]{}%
\providecommand \@@endlink[0]{}%
\providecommand \url  [0]{\begingroup\@sanitize@url \@url }%
\providecommand \@url [1]{\endgroup\@href {#1}{\urlprefix }}%
\providecommand \urlprefix  [0]{URL }%
\providecommand \Eprint [0]{\href }%
\providecommand \doibase [0]{https://doi.org/}%
\providecommand \selectlanguage [0]{\@gobble}%
\providecommand \bibinfo  [0]{\@secondoftwo}%
\providecommand \bibfield  [0]{\@secondoftwo}%
\providecommand \translation [1]{[#1]}%
\providecommand \BibitemOpen [0]{}%
\providecommand \bibitemStop [0]{}%
\providecommand \bibitemNoStop [0]{.\EOS\space}%
\providecommand \EOS [0]{\spacefactor3000\relax}%
\providecommand \BibitemShut  [1]{\csname bibitem#1\endcsname}%
\let\auto@bib@innerbib\@empty
\bibitem [{\citenamefont {Kitaev}(2006)}]{Kitaev2006}%
  \BibitemOpen
  \bibfield  {author} {\bibinfo {author} {\bibfnamefont {A.}~\bibnamefont
  {Kitaev}},\ }\href {https://doi.org/10.1016/j.aop.2005.10.005} {\bibfield
  {journal} {\bibinfo  {journal} {Ann. Phys.}\ }\textbf {\bibinfo {volume}
  {321}},\ \bibinfo {pages} {2} (\bibinfo {year} {2006})}\BibitemShut {NoStop}%
\bibitem [{\citenamefont {Jackeli}\ and\ \citenamefont
  {Khaliullin}(2009)}]{Jackeli2009}%
  \BibitemOpen
  \bibfield  {author} {\bibinfo {author} {\bibfnamefont {G.}~\bibnamefont
  {Jackeli}}\ and\ \bibinfo {author} {\bibfnamefont {G.}~\bibnamefont
  {Khaliullin}},\ }\href {https://doi.org/10.1103/PhysRevLett.102.017205}
  {\bibfield  {journal} {\bibinfo  {journal} {Phys. Rev. Lett.}\ }\textbf
  {\bibinfo {volume} {102}},\ \bibinfo {pages} {017205} (\bibinfo {year}
  {2009})}\BibitemShut {NoStop}%
\bibitem [{\citenamefont {Plumb}\ \emph {et~al.}(2014)\citenamefont {Plumb},
  \citenamefont {Clancy}, \citenamefont {Sandilands}, \citenamefont {Shankar},
  \citenamefont {Hu}, \citenamefont {Burch}, \citenamefont {Kee},\ and\
  \citenamefont {Kim}}]{Plumb2014}%
  \BibitemOpen
  \bibfield  {author} {\bibinfo {author} {\bibfnamefont {K.~W.}\ \bibnamefont
  {Plumb}}, \bibinfo {author} {\bibfnamefont {J.~P.}\ \bibnamefont {Clancy}},
  \bibinfo {author} {\bibfnamefont {L.~J.}\ \bibnamefont {Sandilands}},
  \bibinfo {author} {\bibfnamefont {V.~V.}\ \bibnamefont {Shankar}}, \bibinfo
  {author} {\bibfnamefont {Y.~F.}\ \bibnamefont {Hu}}, \bibinfo {author}
  {\bibfnamefont {K.~S.}\ \bibnamefont {Burch}}, \bibinfo {author}
  {\bibfnamefont {H.-Y.}\ \bibnamefont {Kee}},\ and\ \bibinfo {author}
  {\bibfnamefont {Y.-J.}\ \bibnamefont {Kim}},\ }\href
  {https://doi.org/10.1103/PhysRevB.90.041112} {\bibfield  {journal} {\bibinfo
  {journal} {Phys. Rev. B}\ }\textbf {\bibinfo {volume} {90}},\ \bibinfo
  {pages} {041112} (\bibinfo {year} {2014})}\BibitemShut {NoStop}%
\bibitem [{\citenamefont {Yamaji}\ \emph {et~al.}(2014)\citenamefont {Yamaji},
  \citenamefont {Nomura}, \citenamefont {Kurita}, \citenamefont {Arita},\ and\
  \citenamefont {Imada}}]{Yamaji2014}%
  \BibitemOpen
  \bibfield  {author} {\bibinfo {author} {\bibfnamefont {Y.}~\bibnamefont
  {Yamaji}}, \bibinfo {author} {\bibfnamefont {Y.}~\bibnamefont {Nomura}},
  \bibinfo {author} {\bibfnamefont {M.}~\bibnamefont {Kurita}}, \bibinfo
  {author} {\bibfnamefont {R.}~\bibnamefont {Arita}},\ and\ \bibinfo {author}
  {\bibfnamefont {M.}~\bibnamefont {Imada}},\ }\href
  {https://doi.org/10.1103/PhysRevLett.113.107201} {\bibfield  {journal}
  {\bibinfo  {journal} {Phys. Rev. Lett.}\ }\textbf {\bibinfo {volume} {113}},\
  \bibinfo {pages} {107201} (\bibinfo {year} {2014})}\BibitemShut {NoStop}%
\bibitem [{\citenamefont {Yadav}\ \emph {et~al.}(2016)\citenamefont {Yadav},
  \citenamefont {Bogdanov}, \citenamefont {Katukuri}, \citenamefont
  {Nishimoto}, \citenamefont {van~den Brink},\ and\ \citenamefont
  {Hozoi}}]{Yadav2016}%
  \BibitemOpen
  \bibfield  {author} {\bibinfo {author} {\bibfnamefont {R.}~\bibnamefont
  {Yadav}}, \bibinfo {author} {\bibfnamefont {N.~A.}\ \bibnamefont {Bogdanov}},
  \bibinfo {author} {\bibfnamefont {V.~M.}\ \bibnamefont {Katukuri}}, \bibinfo
  {author} {\bibfnamefont {S.}~\bibnamefont {Nishimoto}}, \bibinfo {author}
  {\bibfnamefont {J.}~\bibnamefont {van~den Brink}},\ and\ \bibinfo {author}
  {\bibfnamefont {L.}~\bibnamefont {Hozoi}},\ }\href
  {https://doi.org/10.1038/srep37925} {\bibfield  {journal} {\bibinfo
  {journal} {Sci. Rep.}\ }\textbf {\bibinfo {volume} {6}},\ \bibinfo {pages}
  {37925} (\bibinfo {year} {2016})}\BibitemShut {NoStop}%
\bibitem [{\citenamefont {Banerjee}\ \emph {et~al.}(2017)\citenamefont
  {Banerjee}, \citenamefont {Yan}, \citenamefont {Knolle}, \citenamefont
  {Bridges}, \citenamefont {Stone}, \citenamefont {Lumsden}, \citenamefont
  {Mandrus}, \citenamefont {Tennant}, \citenamefont {Moessner},\ and\
  \citenamefont {Nagler}}]{Banerjee2017}%
  \BibitemOpen
  \bibfield  {author} {\bibinfo {author} {\bibfnamefont {A.}~\bibnamefont
  {Banerjee}}, \bibinfo {author} {\bibfnamefont {J.}~\bibnamefont {Yan}},
  \bibinfo {author} {\bibfnamefont {J.}~\bibnamefont {Knolle}}, \bibinfo
  {author} {\bibfnamefont {C.~A.}\ \bibnamefont {Bridges}}, \bibinfo {author}
  {\bibfnamefont {M.~B.}\ \bibnamefont {Stone}}, \bibinfo {author}
  {\bibfnamefont {M.~D.}\ \bibnamefont {Lumsden}}, \bibinfo {author}
  {\bibfnamefont {D.~G.}\ \bibnamefont {Mandrus}}, \bibinfo {author}
  {\bibfnamefont {D.~A.}\ \bibnamefont {Tennant}}, \bibinfo {author}
  {\bibfnamefont {R.}~\bibnamefont {Moessner}},\ and\ \bibinfo {author}
  {\bibfnamefont {S.~E.}\ \bibnamefont {Nagler}},\ }\href
  {https://doi.org/10.1126/science.aah6015} {\bibfield  {journal} {\bibinfo
  {journal} {Science}\ }\textbf {\bibinfo {volume} {356}},\ \bibinfo {pages}
  {1055} (\bibinfo {year} {2017})}\BibitemShut {NoStop}%
\bibitem [{\citenamefont {Do}\ \emph {et~al.}(2017)\citenamefont {Do},
  \citenamefont {Park}, \citenamefont {Yoshitake}, \citenamefont {Nasu},
  \citenamefont {Motome}, \citenamefont {Kwon}, \citenamefont {Adroja},
  \citenamefont {Voneshen}, \citenamefont {Kim}, \citenamefont {Jang},
  \citenamefont {Park}, \citenamefont {Choi},\ and\ \citenamefont
  {Ji}}]{Do2017}%
  \BibitemOpen
  \bibfield  {author} {\bibinfo {author} {\bibfnamefont {S.-H.}\ \bibnamefont
  {Do}}, \bibinfo {author} {\bibfnamefont {S.-Y.}\ \bibnamefont {Park}},
  \bibinfo {author} {\bibfnamefont {J.}~\bibnamefont {Yoshitake}}, \bibinfo
  {author} {\bibfnamefont {J.}~\bibnamefont {Nasu}}, \bibinfo {author}
  {\bibfnamefont {Y.}~\bibnamefont {Motome}}, \bibinfo {author} {\bibfnamefont
  {Y.}~\bibnamefont {Kwon}}, \bibinfo {author} {\bibfnamefont {D.~T.}\
  \bibnamefont {Adroja}}, \bibinfo {author} {\bibfnamefont {D.~J.}\
  \bibnamefont {Voneshen}}, \bibinfo {author} {\bibfnamefont {K.}~\bibnamefont
  {Kim}}, \bibinfo {author} {\bibfnamefont {T.-H.}\ \bibnamefont {Jang}},
  \bibinfo {author} {\bibfnamefont {J.-H.}\ \bibnamefont {Park}}, \bibinfo
  {author} {\bibfnamefont {K.-Y.}\ \bibnamefont {Choi}},\ and\ \bibinfo
  {author} {\bibfnamefont {S.}~\bibnamefont {Ji}},\ }\href
  {https://doi.org/10.1038/nphys4264} {\bibfield  {journal} {\bibinfo
  {journal} {Nat. Phys.}\ }\textbf {\bibinfo {volume} {13}},\ \bibinfo {pages}
  {1079} (\bibinfo {year} {2017})}\BibitemShut {NoStop}%
\bibitem [{\citenamefont {Kim}\ \emph {et~al.}(2020)\citenamefont {Kim},
  \citenamefont {Chaloupka}, \citenamefont {Singh}, \citenamefont {Kim},
  \citenamefont {Kim}, \citenamefont {Casa}, \citenamefont {Said},
  \citenamefont {Huang},\ and\ \citenamefont {Gog}}]{Kim2020}%
  \BibitemOpen
  \bibfield  {author} {\bibinfo {author} {\bibfnamefont {J.}~\bibnamefont
  {Kim}}, \bibinfo {author} {\bibfnamefont {J.~c.~v.}\ \bibnamefont
  {Chaloupka}}, \bibinfo {author} {\bibfnamefont {Y.}~\bibnamefont {Singh}},
  \bibinfo {author} {\bibfnamefont {J.~W.}\ \bibnamefont {Kim}}, \bibinfo
  {author} {\bibfnamefont {B.~J.}\ \bibnamefont {Kim}}, \bibinfo {author}
  {\bibfnamefont {D.}~\bibnamefont {Casa}}, \bibinfo {author} {\bibfnamefont
  {A.}~\bibnamefont {Said}}, \bibinfo {author} {\bibfnamefont {X.}~\bibnamefont
  {Huang}},\ and\ \bibinfo {author} {\bibfnamefont {T.}~\bibnamefont {Gog}},\
  }\href {https://doi.org/10.1103/PhysRevX.10.021034} {\bibfield  {journal}
  {\bibinfo  {journal} {Phys. Rev. X}\ }\textbf {\bibinfo {volume} {10}},\
  \bibinfo {pages} {021034} (\bibinfo {year} {2020})}\BibitemShut {NoStop}%
\bibitem [{\citenamefont {Rau}\ \emph {et~al.}(2016)\citenamefont {Rau},
  \citenamefont {Lee},\ and\ \citenamefont {Kee}}]{Rau2016}%
  \BibitemOpen
  \bibfield  {author} {\bibinfo {author} {\bibfnamefont {J.~G.}\ \bibnamefont
  {Rau}}, \bibinfo {author} {\bibfnamefont {E.~K.-H.}\ \bibnamefont {Lee}},\
  and\ \bibinfo {author} {\bibfnamefont {H.-Y.}\ \bibnamefont {Kee}},\ }\href
  {https://doi.org/10.1146/annurev-conmatphys-031115-011319} {\bibfield
  {journal} {\bibinfo  {journal} {Ann. Rev. Condens. Matter}\ }\textbf
  {\bibinfo {volume} {7}},\ \bibinfo {pages} {195} (\bibinfo {year}
  {2016})}\BibitemShut {NoStop}%
\bibitem [{\citenamefont {Hermanns}\ \emph {et~al.}(2018)\citenamefont
  {Hermanns}, \citenamefont {Kimchi},\ and\ \citenamefont
  {Knolle}}]{Hermanns2018}%
  \BibitemOpen
  \bibfield  {author} {\bibinfo {author} {\bibfnamefont {M.}~\bibnamefont
  {Hermanns}}, \bibinfo {author} {\bibfnamefont {I.}~\bibnamefont {Kimchi}},\
  and\ \bibinfo {author} {\bibfnamefont {J.}~\bibnamefont {Knolle}},\ }\href
  {https://doi.org/10.1146/annurev-conmatphys-033117-053934} {\bibfield
  {journal} {\bibinfo  {journal} {Annu. Rev. Condens. Matter}\ }\textbf
  {\bibinfo {volume} {9}},\ \bibinfo {pages} {17} (\bibinfo {year}
  {2018})}\BibitemShut {NoStop}%
\bibitem [{\citenamefont {Takagi}\ \emph {et~al.}(2019)\citenamefont {Takagi},
  \citenamefont {Takayama}, \citenamefont {Jackeli}, \citenamefont
  {Khaliullin},\ and\ \citenamefont {Nagler}}]{Takagi2019}%
  \BibitemOpen
  \bibfield  {author} {\bibinfo {author} {\bibfnamefont {H.}~\bibnamefont
  {Takagi}}, \bibinfo {author} {\bibfnamefont {T.}~\bibnamefont {Takayama}},
  \bibinfo {author} {\bibfnamefont {G.}~\bibnamefont {Jackeli}}, \bibinfo
  {author} {\bibfnamefont {G.}~\bibnamefont {Khaliullin}},\ and\ \bibinfo
  {author} {\bibfnamefont {S.~E.}\ \bibnamefont {Nagler}},\ }\href
  {https://doi.org/10.1038/s42254-019-0038-2} {\bibfield  {journal} {\bibinfo
  {journal} {Nat. Rev. Phys.}\ }\textbf {\bibinfo {volume} {1}},\ \bibinfo
  {pages} {264} (\bibinfo {year} {2019})}\BibitemShut {NoStop}%
\bibitem [{\citenamefont {Khomskii}\ and\ \citenamefont
  {Streltsov}(2021)}]{Khomskii2021}%
  \BibitemOpen
  \bibfield  {author} {\bibinfo {author} {\bibfnamefont {D.~I.}\ \bibnamefont
  {Khomskii}}\ and\ \bibinfo {author} {\bibfnamefont {S.~V.}\ \bibnamefont
  {Streltsov}},\ }\href {https://doi.org/10.1021/acs.chemrev.0c00579}
  {\bibfield  {journal} {\bibinfo  {journal} {Chem. Rev.}\ }\textbf {\bibinfo
  {volume} {121}},\ \bibinfo {pages} {2992} (\bibinfo {year}
  {2021})}\BibitemShut {NoStop}%
\bibitem [{\citenamefont {Singh}\ \emph {et~al.}(2012)\citenamefont {Singh},
  \citenamefont {Manni}, \citenamefont {Reuther}, \citenamefont {Berlijn},
  \citenamefont {Thomale}, \citenamefont {Ku}, \citenamefont {Trebst},\ and\
  \citenamefont {Gegenwart}}]{Singh2012}%
  \BibitemOpen
  \bibfield  {author} {\bibinfo {author} {\bibfnamefont {Y.}~\bibnamefont
  {Singh}}, \bibinfo {author} {\bibfnamefont {S.}~\bibnamefont {Manni}},
  \bibinfo {author} {\bibfnamefont {J.}~\bibnamefont {Reuther}}, \bibinfo
  {author} {\bibfnamefont {T.}~\bibnamefont {Berlijn}}, \bibinfo {author}
  {\bibfnamefont {R.}~\bibnamefont {Thomale}}, \bibinfo {author} {\bibfnamefont
  {W.}~\bibnamefont {Ku}}, \bibinfo {author} {\bibfnamefont {S.}~\bibnamefont
  {Trebst}},\ and\ \bibinfo {author} {\bibfnamefont {P.}~\bibnamefont
  {Gegenwart}},\ }\href {https://doi.org/10.1103/PhysRevLett.108.127203}
  {\bibfield  {journal} {\bibinfo  {journal} {Phys. Rev. Lett.}\ }\textbf
  {\bibinfo {volume} {108}},\ \bibinfo {pages} {127203} (\bibinfo {year}
  {2012})}\BibitemShut {NoStop}%
\bibitem [{\citenamefont {Chaloupka}\ \emph {et~al.}(2010)\citenamefont
  {Chaloupka}, \citenamefont {Jackeli},\ and\ \citenamefont
  {Khaliullin}}]{Chaloupka2010}%
  \BibitemOpen
  \bibfield  {author} {\bibinfo {author} {\bibfnamefont {J.}~\bibnamefont
  {Chaloupka}}, \bibinfo {author} {\bibfnamefont {G.}~\bibnamefont {Jackeli}},\
  and\ \bibinfo {author} {\bibfnamefont {G.}~\bibnamefont {Khaliullin}},\
  }\href {https://doi.org/10.1103/PhysRevLett.105.027204} {\bibfield  {journal}
  {\bibinfo  {journal} {Phys. Rev. Lett.}\ }\textbf {\bibinfo {volume} {105}},\
  \bibinfo {pages} {027204} (\bibinfo {year} {2010})}\BibitemShut {NoStop}%
\bibitem [{\citenamefont {Sizyuk}\ \emph {et~al.}(2014)\citenamefont {Sizyuk},
  \citenamefont {Price}, \citenamefont {W\"olfle},\ and\ \citenamefont
  {Perkins}}]{Sizyuk2014}%
  \BibitemOpen
  \bibfield  {author} {\bibinfo {author} {\bibfnamefont {Y.}~\bibnamefont
  {Sizyuk}}, \bibinfo {author} {\bibfnamefont {C.}~\bibnamefont {Price}},
  \bibinfo {author} {\bibfnamefont {P.}~\bibnamefont {W\"olfle}},\ and\
  \bibinfo {author} {\bibfnamefont {N.~B.}\ \bibnamefont {Perkins}},\ }\href
  {https://doi.org/10.1103/PhysRevB.90.155126} {\bibfield  {journal} {\bibinfo
  {journal} {Phys. Rev. B}\ }\textbf {\bibinfo {volume} {90}},\ \bibinfo
  {pages} {155126} (\bibinfo {year} {2014})}\BibitemShut {NoStop}%
\bibitem [{\citenamefont {Rau}\ \emph {et~al.}(2014)\citenamefont {Rau},
  \citenamefont {Lee},\ and\ \citenamefont {Kee}}]{Rau2014}%
  \BibitemOpen
  \bibfield  {author} {\bibinfo {author} {\bibfnamefont {J.~G.}\ \bibnamefont
  {Rau}}, \bibinfo {author} {\bibfnamefont {E.~K.-H.}\ \bibnamefont {Lee}},\
  and\ \bibinfo {author} {\bibfnamefont {H.-Y.}\ \bibnamefont {Kee}},\ }\href
  {https://doi.org/10.1103/PhysRevLett.112.077204} {\bibfield  {journal}
  {\bibinfo  {journal} {Phys. Rev. Lett.}\ }\textbf {\bibinfo {volume} {112}},\
  \bibinfo {pages} {077204} (\bibinfo {year} {2014})}\BibitemShut {NoStop}%
\bibitem [{\citenamefont {Rau}\ and\ \citenamefont {Kee}(2014)}]{Rau2014_2}%
  \BibitemOpen
  \bibfield  {author} {\bibinfo {author} {\bibfnamefont {J.~G.}\ \bibnamefont
  {Rau}}\ and\ \bibinfo {author} {\bibfnamefont {H.-Y.}\ \bibnamefont {Kee}},\
  }\href {http://arxiv.org/abs/1408.4811} {\bibfield  {journal} {\bibinfo
  {journal} {arXiv:1408.4811}\ } (\bibinfo {year} {2014})}\BibitemShut
  {NoStop}%
\bibitem [{\citenamefont {Sears}\ \emph {et~al.}(2015)\citenamefont {Sears},
  \citenamefont {Songvilay}, \citenamefont {Plumb}, \citenamefont {Clancy},
  \citenamefont {Qiu}, \citenamefont {Zhao}, \citenamefont {Parshall},\ and\
  \citenamefont {Kim}}]{Sears2015}%
  \BibitemOpen
  \bibfield  {author} {\bibinfo {author} {\bibfnamefont {J.~A.}\ \bibnamefont
  {Sears}}, \bibinfo {author} {\bibfnamefont {M.}~\bibnamefont {Songvilay}},
  \bibinfo {author} {\bibfnamefont {K.~W.}\ \bibnamefont {Plumb}}, \bibinfo
  {author} {\bibfnamefont {J.~P.}\ \bibnamefont {Clancy}}, \bibinfo {author}
  {\bibfnamefont {Y.}~\bibnamefont {Qiu}}, \bibinfo {author} {\bibfnamefont
  {Y.}~\bibnamefont {Zhao}}, \bibinfo {author} {\bibfnamefont {D.}~\bibnamefont
  {Parshall}},\ and\ \bibinfo {author} {\bibfnamefont {Y.-J.}\ \bibnamefont
  {Kim}},\ }\href {https://doi.org/10.1103/PhysRevB.91.144420} {\bibfield
  {journal} {\bibinfo  {journal} {Phys. Rev. B}\ }\textbf {\bibinfo {volume}
  {91}},\ \bibinfo {pages} {144420} (\bibinfo {year} {2015})}\BibitemShut
  {NoStop}%
\bibitem [{\citenamefont {Chaloupka}\ and\ \citenamefont
  {Khaliullin}(2016)}]{Chaloupka2016}%
  \BibitemOpen
  \bibfield  {author} {\bibinfo {author} {\bibfnamefont {J.}~\bibnamefont
  {Chaloupka}}\ and\ \bibinfo {author} {\bibfnamefont {G.}~\bibnamefont
  {Khaliullin}},\ }\href {https://doi.org/10.1103/PhysRevB.94.064435}
  {\bibfield  {journal} {\bibinfo  {journal} {Phys. Rev. B}\ }\textbf {\bibinfo
  {volume} {94}},\ \bibinfo {pages} {064435} (\bibinfo {year}
  {2016})}\BibitemShut {NoStop}%
\bibitem [{\citenamefont {Sizyuk}\ \emph {et~al.}(2016)\citenamefont {Sizyuk},
  \citenamefont {W\"olfle},\ and\ \citenamefont {Perkins}}]{Sizyuk2016}%
  \BibitemOpen
  \bibfield  {author} {\bibinfo {author} {\bibfnamefont {Y.}~\bibnamefont
  {Sizyuk}}, \bibinfo {author} {\bibfnamefont {P.}~\bibnamefont {W\"olfle}},\
  and\ \bibinfo {author} {\bibfnamefont {N.~B.}\ \bibnamefont {Perkins}},\
  }\href {https://doi.org/10.1103/PhysRevB.94.085109} {\bibfield  {journal}
  {\bibinfo  {journal} {Phys. Rev. B}\ }\textbf {\bibinfo {volume} {94}},\
  \bibinfo {pages} {085109} (\bibinfo {year} {2016})}\BibitemShut {NoStop}%
\bibitem [{\citenamefont {Rousochatzakis}\ \emph {et~al.}(2018)\citenamefont
  {Rousochatzakis}, \citenamefont {Sizyuk},\ and\ \citenamefont
  {Perkins}}]{Rousochatzakis2018}%
  \BibitemOpen
  \bibfield  {author} {\bibinfo {author} {\bibfnamefont {I.}~\bibnamefont
  {Rousochatzakis}}, \bibinfo {author} {\bibfnamefont {Y.}~\bibnamefont
  {Sizyuk}},\ and\ \bibinfo {author} {\bibfnamefont {N.~B.}\ \bibnamefont
  {Perkins}},\ }\href {https://doi.org/10.1038/s41467-018-03934-1} {\bibfield
  {journal} {\bibinfo  {journal} {Nat. Commun.}\ }\textbf {\bibinfo {volume}
  {9}},\ \bibinfo {pages} {1575} (\bibinfo {year} {2018})}\BibitemShut
  {NoStop}%
\bibitem [{\citenamefont {Tovar-Olvera}\ \emph {et~al.}(2022)\citenamefont
  {Tovar-Olvera}, \citenamefont {Ruiz-D\'{\i}az},\ and\ \citenamefont
  {Sauban\`ere}}]{Tovar2022}%
  \BibitemOpen
  \bibfield  {author} {\bibinfo {author} {\bibfnamefont {M.~A.}\ \bibnamefont
  {Tovar-Olvera}}, \bibinfo {author} {\bibfnamefont {P.}~\bibnamefont
  {Ruiz-D\'{\i}az}},\ and\ \bibinfo {author} {\bibfnamefont {M.}~\bibnamefont
  {Sauban\`ere}},\ }\href {https://doi.org/10.1103/PhysRevB.105.094413}
  {\bibfield  {journal} {\bibinfo  {journal} {Phys. Rev. B}\ }\textbf {\bibinfo
  {volume} {105}},\ \bibinfo {pages} {094413} (\bibinfo {year}
  {2022})}\BibitemShut {NoStop}%
\bibitem [{\citenamefont {Winter}\ \emph {et~al.}(2016)\citenamefont {Winter},
  \citenamefont {Li}, \citenamefont {Jeschke},\ and\ \citenamefont
  {Valent\'{\i}}}]{Winter2016}%
  \BibitemOpen
  \bibfield  {author} {\bibinfo {author} {\bibfnamefont {S.~M.}\ \bibnamefont
  {Winter}}, \bibinfo {author} {\bibfnamefont {Y.}~\bibnamefont {Li}}, \bibinfo
  {author} {\bibfnamefont {H.~O.}\ \bibnamefont {Jeschke}},\ and\ \bibinfo
  {author} {\bibfnamefont {R.}~\bibnamefont {Valent\'{\i}}},\ }\href
  {https://doi.org/10.1103/PhysRevB.93.214431} {\bibfield  {journal} {\bibinfo
  {journal} {Phys. Rev. B}\ }\textbf {\bibinfo {volume} {93}},\ \bibinfo
  {pages} {214431} (\bibinfo {year} {2016})}\BibitemShut {NoStop}%
\bibitem [{\citenamefont {Winter}\ \emph {et~al.}(2017)\citenamefont {Winter},
  \citenamefont {Tsirlin}, \citenamefont {Daghofer}, \citenamefont {van~den
  Brink}, \citenamefont {Singh}, \citenamefont {Gegenwart},\ and\ \citenamefont
  {Valentí}}]{Winter2017}%
  \BibitemOpen
  \bibfield  {author} {\bibinfo {author} {\bibfnamefont {S.~M.}\ \bibnamefont
  {Winter}}, \bibinfo {author} {\bibfnamefont {A.~A.}\ \bibnamefont {Tsirlin}},
  \bibinfo {author} {\bibfnamefont {M.}~\bibnamefont {Daghofer}}, \bibinfo
  {author} {\bibfnamefont {J.}~\bibnamefont {van~den Brink}}, \bibinfo {author}
  {\bibfnamefont {Y.}~\bibnamefont {Singh}}, \bibinfo {author} {\bibfnamefont
  {P.}~\bibnamefont {Gegenwart}},\ and\ \bibinfo {author} {\bibfnamefont
  {R.}~\bibnamefont {Valentí}},\ }\href
  {https://doi.org/10.1088/1361-648X/aa8cf5} {\bibfield  {journal} {\bibinfo
  {journal} {J. Phys. Conden. Matter}\ }\textbf {\bibinfo {volume} {29}},\
  \bibinfo {pages} {493002} (\bibinfo {year} {2017})}\BibitemShut {NoStop}%
\bibitem [{\citenamefont {Sears}\ \emph {et~al.}(2020)\citenamefont {Sears},
  \citenamefont {Chern}, \citenamefont {Kim}, \citenamefont {Bereciartua},
  \citenamefont {Francoual}, \citenamefont {Kim},\ and\ \citenamefont
  {Kim}}]{Sears2020}%
  \BibitemOpen
  \bibfield  {author} {\bibinfo {author} {\bibfnamefont {J.~A.}\ \bibnamefont
  {Sears}}, \bibinfo {author} {\bibfnamefont {L.~E.}\ \bibnamefont {Chern}},
  \bibinfo {author} {\bibfnamefont {S.}~\bibnamefont {Kim}}, \bibinfo {author}
  {\bibfnamefont {P.~J.}\ \bibnamefont {Bereciartua}}, \bibinfo {author}
  {\bibfnamefont {S.}~\bibnamefont {Francoual}}, \bibinfo {author}
  {\bibfnamefont {Y.~B.}\ \bibnamefont {Kim}},\ and\ \bibinfo {author}
  {\bibfnamefont {Y.-J.}\ \bibnamefont {Kim}},\ }\href
  {https://doi.org/10.1038/s41567-020-0874-0} {\bibfield  {journal} {\bibinfo
  {journal} {Nat. Phys.}\ }\textbf {\bibinfo {volume} {16}},\ \bibinfo {pages}
  {837} (\bibinfo {year} {2020})}\BibitemShut {NoStop}%
\bibitem [{\citenamefont {Suzuki}\ \emph {et~al.}(2021)\citenamefont {Suzuki},
  \citenamefont {Liu}, \citenamefont {Bertinshaw}, \citenamefont {Ueda},
  \citenamefont {Kim}, \citenamefont {Laha}, \citenamefont {Weber},
  \citenamefont {Yang}, \citenamefont {Wang}, \citenamefont {Takahashi},
  \citenamefont {Fürsich}, \citenamefont {Minola}, \citenamefont {Lotsch},
  \citenamefont {Kim}, \citenamefont {Yavaş}, \citenamefont {Daghofer},
  \citenamefont {Chaloupka}, \citenamefont {Khaliullin}, \citenamefont
  {Gretarsson},\ and\ \citenamefont {Keimer}}]{Suzuki2021}%
  \BibitemOpen
  \bibfield  {author} {\bibinfo {author} {\bibfnamefont {H.}~\bibnamefont
  {Suzuki}}, \bibinfo {author} {\bibfnamefont {H.}~\bibnamefont {Liu}},
  \bibinfo {author} {\bibfnamefont {J.}~\bibnamefont {Bertinshaw}}, \bibinfo
  {author} {\bibfnamefont {K.}~\bibnamefont {Ueda}}, \bibinfo {author}
  {\bibfnamefont {H.}~\bibnamefont {Kim}}, \bibinfo {author} {\bibfnamefont
  {S.}~\bibnamefont {Laha}}, \bibinfo {author} {\bibfnamefont {D.}~\bibnamefont
  {Weber}}, \bibinfo {author} {\bibfnamefont {Z.}~\bibnamefont {Yang}},
  \bibinfo {author} {\bibfnamefont {L.}~\bibnamefont {Wang}}, \bibinfo {author}
  {\bibfnamefont {H.}~\bibnamefont {Takahashi}}, \bibinfo {author}
  {\bibfnamefont {K.}~\bibnamefont {Fürsich}}, \bibinfo {author}
  {\bibfnamefont {M.}~\bibnamefont {Minola}}, \bibinfo {author} {\bibfnamefont
  {B.~V.}\ \bibnamefont {Lotsch}}, \bibinfo {author} {\bibfnamefont {B.~J.}\
  \bibnamefont {Kim}}, \bibinfo {author} {\bibfnamefont {H.}~\bibnamefont
  {Yavaş}}, \bibinfo {author} {\bibfnamefont {M.}~\bibnamefont {Daghofer}},
  \bibinfo {author} {\bibfnamefont {J.}~\bibnamefont {Chaloupka}}, \bibinfo
  {author} {\bibfnamefont {G.}~\bibnamefont {Khaliullin}}, \bibinfo {author}
  {\bibfnamefont {H.}~\bibnamefont {Gretarsson}},\ and\ \bibinfo {author}
  {\bibfnamefont {B.}~\bibnamefont {Keimer}},\ }\href
  {https://doi.org/10.1038/s41467-021-24722-4} {\bibfield  {journal} {\bibinfo
  {journal} {Nat. Commun.}\ }\textbf {\bibinfo {volume} {12}},\ \bibinfo
  {pages} {4512} (\bibinfo {year} {2021})}\BibitemShut {NoStop}%
\bibitem [{\citenamefont {Sears}\ \emph {et~al.}(2017)\citenamefont {Sears},
  \citenamefont {Zhao}, \citenamefont {Xu}, \citenamefont {Lynn},\ and\
  \citenamefont {Kim}}]{Sears2017}%
  \BibitemOpen
  \bibfield  {author} {\bibinfo {author} {\bibfnamefont {J.~A.}\ \bibnamefont
  {Sears}}, \bibinfo {author} {\bibfnamefont {Y.}~\bibnamefont {Zhao}},
  \bibinfo {author} {\bibfnamefont {Z.}~\bibnamefont {Xu}}, \bibinfo {author}
  {\bibfnamefont {J.~W.}\ \bibnamefont {Lynn}},\ and\ \bibinfo {author}
  {\bibfnamefont {Y.-J.}\ \bibnamefont {Kim}},\ }\href
  {https://doi.org/10.1103/PhysRevB.95.180411} {\bibfield  {journal} {\bibinfo
  {journal} {Phys. Rev. B}\ }\textbf {\bibinfo {volume} {95}},\ \bibinfo
  {pages} {180411} (\bibinfo {year} {2017})}\BibitemShut {NoStop}%
\bibitem [{\citenamefont {Baek}\ \emph {et~al.}(2017)\citenamefont {Baek},
  \citenamefont {Do}, \citenamefont {Choi}, \citenamefont {Kwon}, \citenamefont
  {Wolter}, \citenamefont {Nishimoto}, \citenamefont {van~den Brink},\ and\
  \citenamefont {B\"uchner}}]{Baek2017}%
  \BibitemOpen
  \bibfield  {author} {\bibinfo {author} {\bibfnamefont {S.-H.}\ \bibnamefont
  {Baek}}, \bibinfo {author} {\bibfnamefont {S.-H.}\ \bibnamefont {Do}},
  \bibinfo {author} {\bibfnamefont {K.-Y.}\ \bibnamefont {Choi}}, \bibinfo
  {author} {\bibfnamefont {Y.~S.}\ \bibnamefont {Kwon}}, \bibinfo {author}
  {\bibfnamefont {A.~U.~B.}\ \bibnamefont {Wolter}}, \bibinfo {author}
  {\bibfnamefont {S.}~\bibnamefont {Nishimoto}}, \bibinfo {author}
  {\bibfnamefont {J.}~\bibnamefont {van~den Brink}},\ and\ \bibinfo {author}
  {\bibfnamefont {B.}~\bibnamefont {B\"uchner}},\ }\href
  {https://doi.org/10.1103/PhysRevLett.119.037201} {\bibfield  {journal}
  {\bibinfo  {journal} {Phys. Rev. Lett.}\ }\textbf {\bibinfo {volume} {119}},\
  \bibinfo {pages} {037201} (\bibinfo {year} {2017})}\BibitemShut {NoStop}%
\bibitem [{\citenamefont {Kasahara}\ \emph {et~al.}(2018)\citenamefont
  {Kasahara}, \citenamefont {Ohnishi}, \citenamefont {Mizukami}, \citenamefont
  {Tanaka}, \citenamefont {Ma}, \citenamefont {Sugii}, \citenamefont {Kurita},
  \citenamefont {Tanaka}, \citenamefont {Nasu}, \citenamefont {Motome},
  \citenamefont {Shibauchi},\ and\ \citenamefont {Matsuda}}]{Kasahara2018}%
  \BibitemOpen
  \bibfield  {author} {\bibinfo {author} {\bibfnamefont {Y.}~\bibnamefont
  {Kasahara}}, \bibinfo {author} {\bibfnamefont {T.}~\bibnamefont {Ohnishi}},
  \bibinfo {author} {\bibfnamefont {Y.}~\bibnamefont {Mizukami}}, \bibinfo
  {author} {\bibfnamefont {O.}~\bibnamefont {Tanaka}}, \bibinfo {author}
  {\bibfnamefont {S.}~\bibnamefont {Ma}}, \bibinfo {author} {\bibfnamefont
  {K.}~\bibnamefont {Sugii}}, \bibinfo {author} {\bibfnamefont
  {N.}~\bibnamefont {Kurita}}, \bibinfo {author} {\bibfnamefont
  {H.}~\bibnamefont {Tanaka}}, \bibinfo {author} {\bibfnamefont
  {J.}~\bibnamefont {Nasu}}, \bibinfo {author} {\bibfnamefont {Y.}~\bibnamefont
  {Motome}}, \bibinfo {author} {\bibfnamefont {T.}~\bibnamefont {Shibauchi}},\
  and\ \bibinfo {author} {\bibfnamefont {Y.}~\bibnamefont {Matsuda}},\ }\href
  {https://doi.org/10.1038/s41586-018-0274-0} {\bibfield  {journal} {\bibinfo
  {journal} {Nature}\ }\textbf {\bibinfo {volume} {559}},\ \bibinfo {pages}
  {227} (\bibinfo {year} {2018})}\BibitemShut {NoStop}%
\bibitem [{\citenamefont {Czajka}\ \emph {et~al.}(2021)\citenamefont {Czajka},
  \citenamefont {Gao}, \citenamefont {Hirschberger}, \citenamefont
  {Lampen-Kelley}, \citenamefont {Banerjee}, \citenamefont {Yan}, \citenamefont
  {Mandrus}, \citenamefont {Nagler},\ and\ \citenamefont {Ong}}]{Czajka2021}%
  \BibitemOpen
  \bibfield  {author} {\bibinfo {author} {\bibfnamefont {P.}~\bibnamefont
  {Czajka}}, \bibinfo {author} {\bibfnamefont {T.}~\bibnamefont {Gao}},
  \bibinfo {author} {\bibfnamefont {M.}~\bibnamefont {Hirschberger}}, \bibinfo
  {author} {\bibfnamefont {P.}~\bibnamefont {Lampen-Kelley}}, \bibinfo {author}
  {\bibfnamefont {A.}~\bibnamefont {Banerjee}}, \bibinfo {author}
  {\bibfnamefont {J.}~\bibnamefont {Yan}}, \bibinfo {author} {\bibfnamefont
  {D.~G.}\ \bibnamefont {Mandrus}}, \bibinfo {author} {\bibfnamefont {S.~E.}\
  \bibnamefont {Nagler}},\ and\ \bibinfo {author} {\bibfnamefont {N.~P.}\
  \bibnamefont {Ong}},\ }\href {https://doi.org/10.1038/s41567-021-01243-x}
  {\bibfield  {journal} {\bibinfo  {journal} {Nat. Phys.}\ }\textbf {\bibinfo
  {volume} {17}},\ \bibinfo {pages} {915} (\bibinfo {year} {2021})}\BibitemShut
  {NoStop}%
\bibitem [{\citenamefont {Bruin}\ \emph {et~al.}(2022)\citenamefont {Bruin},
  \citenamefont {Claus}, \citenamefont {Matsumoto}, \citenamefont {Kurita},
  \citenamefont {Tanaka},\ and\ \citenamefont {Takagi}}]{Bruin2022}%
  \BibitemOpen
  \bibfield  {author} {\bibinfo {author} {\bibfnamefont {J.~A.~N.}\
  \bibnamefont {Bruin}}, \bibinfo {author} {\bibfnamefont {R.~R.}\ \bibnamefont
  {Claus}}, \bibinfo {author} {\bibfnamefont {Y.}~\bibnamefont {Matsumoto}},
  \bibinfo {author} {\bibfnamefont {N.}~\bibnamefont {Kurita}}, \bibinfo
  {author} {\bibfnamefont {H.}~\bibnamefont {Tanaka}},\ and\ \bibinfo {author}
  {\bibfnamefont {H.}~\bibnamefont {Takagi}},\ }\href
  {https://doi.org/10.1038/s41567-021-01501-y} {\bibfield  {journal} {\bibinfo
  {journal} {Nat. Phys.}\ }\textbf {\bibinfo {volume} {18}},\ \bibinfo {pages}
  {401} (\bibinfo {year} {2022})}\BibitemShut {NoStop}%
\bibitem [{\citenamefont {Tanaka}\ \emph {et~al.}(2022)\citenamefont {Tanaka},
  \citenamefont {Mizukami}, \citenamefont {Harasawa}, \citenamefont
  {Hashimoto}, \citenamefont {Hwang}, \citenamefont {Kurita}, \citenamefont
  {Tanaka}, \citenamefont {Fujimoto}, \citenamefont {Matsuda}, \citenamefont
  {Moon},\ and\ \citenamefont {Shibauchi}}]{Tanaka2022}%
  \BibitemOpen
  \bibfield  {author} {\bibinfo {author} {\bibfnamefont {O.}~\bibnamefont
  {Tanaka}}, \bibinfo {author} {\bibfnamefont {Y.}~\bibnamefont {Mizukami}},
  \bibinfo {author} {\bibfnamefont {R.}~\bibnamefont {Harasawa}}, \bibinfo
  {author} {\bibfnamefont {K.}~\bibnamefont {Hashimoto}}, \bibinfo {author}
  {\bibfnamefont {K.}~\bibnamefont {Hwang}}, \bibinfo {author} {\bibfnamefont
  {N.}~\bibnamefont {Kurita}}, \bibinfo {author} {\bibfnamefont
  {H.}~\bibnamefont {Tanaka}}, \bibinfo {author} {\bibfnamefont
  {S.}~\bibnamefont {Fujimoto}}, \bibinfo {author} {\bibfnamefont
  {Y.}~\bibnamefont {Matsuda}}, \bibinfo {author} {\bibfnamefont {E.~G.}\
  \bibnamefont {Moon}},\ and\ \bibinfo {author} {\bibfnamefont
  {T.}~\bibnamefont {Shibauchi}},\ }\href
  {https://doi.org/10.1038/s41567-021-01488-6} {\bibfield  {journal} {\bibinfo
  {journal} {Nat. Phys.}\ }\textbf {\bibinfo {volume} {18}},\ \bibinfo {pages}
  {429} (\bibinfo {year} {2022})}\BibitemShut {NoStop}%
\bibitem [{\citenamefont {Czajka}\ \emph {et~al.}(2023)\citenamefont {Czajka},
  \citenamefont {Gao}, \citenamefont {Hirschberger}, \citenamefont
  {Lampen-Kelley}, \citenamefont {Banerjee}, \citenamefont {Quirk},
  \citenamefont {Mandrus}, \citenamefont {Nagler},\ and\ \citenamefont
  {Ong}}]{Czajka2023}%
  \BibitemOpen
  \bibfield  {author} {\bibinfo {author} {\bibfnamefont {P.}~\bibnamefont
  {Czajka}}, \bibinfo {author} {\bibfnamefont {T.}~\bibnamefont {Gao}},
  \bibinfo {author} {\bibfnamefont {M.}~\bibnamefont {Hirschberger}}, \bibinfo
  {author} {\bibfnamefont {P.}~\bibnamefont {Lampen-Kelley}}, \bibinfo {author}
  {\bibfnamefont {A.}~\bibnamefont {Banerjee}}, \bibinfo {author}
  {\bibfnamefont {N.}~\bibnamefont {Quirk}}, \bibinfo {author} {\bibfnamefont
  {D.~G.}\ \bibnamefont {Mandrus}}, \bibinfo {author} {\bibfnamefont {S.~E.}\
  \bibnamefont {Nagler}},\ and\ \bibinfo {author} {\bibfnamefont {N.~P.}\
  \bibnamefont {Ong}},\ }\href {https://doi.org/10.1038/s41563-022-01397-w}
  {\bibfield  {journal} {\bibinfo  {journal} {Nat. Mater.}\ }\textbf {\bibinfo
  {volume} {22}},\ \bibinfo {pages} {36} (\bibinfo {year} {2023})}\BibitemShut
  {NoStop}%
\bibitem [{\citenamefont {R\"uegg}\ \emph {et~al.}(2008)\citenamefont
  {R\"uegg}, \citenamefont {Normand}, \citenamefont {Matsumoto}, \citenamefont
  {Furrer}, \citenamefont {McMorrow}, \citenamefont {Kr\"amer}, \citenamefont
  {G\"udel}, \citenamefont {Gvasaliya}, \citenamefont {Mutka},\ and\
  \citenamefont {Boehm}}]{Rueegg2008}%
  \BibitemOpen
  \bibfield  {author} {\bibinfo {author} {\bibfnamefont {C.}~\bibnamefont
  {R\"uegg}}, \bibinfo {author} {\bibfnamefont {B.}~\bibnamefont {Normand}},
  \bibinfo {author} {\bibfnamefont {M.}~\bibnamefont {Matsumoto}}, \bibinfo
  {author} {\bibfnamefont {A.}~\bibnamefont {Furrer}}, \bibinfo {author}
  {\bibfnamefont {D.~F.}\ \bibnamefont {McMorrow}}, \bibinfo {author}
  {\bibfnamefont {K.~W.}\ \bibnamefont {Kr\"amer}}, \bibinfo {author}
  {\bibfnamefont {H.~U.}\ \bibnamefont {G\"udel}}, \bibinfo {author}
  {\bibfnamefont {S.~N.}\ \bibnamefont {Gvasaliya}}, \bibinfo {author}
  {\bibfnamefont {H.}~\bibnamefont {Mutka}},\ and\ \bibinfo {author}
  {\bibfnamefont {M.}~\bibnamefont {Boehm}},\ }\href
  {https://doi.org/10.1103/PhysRevLett.100.205701} {\bibfield  {journal}
  {\bibinfo  {journal} {Phys. Rev. Lett.}\ }\textbf {\bibinfo {volume} {100}},\
  \bibinfo {pages} {205701} (\bibinfo {year} {2008})}\BibitemShut {NoStop}%
\bibitem [{\citenamefont {Gati}\ \emph {et~al.}(2020)\citenamefont {Gati},
  \citenamefont {Xiang}, \citenamefont {Bud'ko},\ and\ \citenamefont
  {Canfield}}]{Gati2020}%
  \BibitemOpen
  \bibfield  {author} {\bibinfo {author} {\bibfnamefont {E.}~\bibnamefont
  {Gati}}, \bibinfo {author} {\bibfnamefont {L.}~\bibnamefont {Xiang}},
  \bibinfo {author} {\bibfnamefont {S.~L.}\ \bibnamefont {Bud'ko}},\ and\
  \bibinfo {author} {\bibfnamefont {P.~C.}\ \bibnamefont {Canfield}},\ }\href
  {https://doi.org/10.1002/andp.202000248} {\bibfield  {journal} {\bibinfo
  {journal} {Ann. Phys.}\ }\textbf {\bibinfo {volume} {532}},\ \bibinfo {pages}
  {2000248} (\bibinfo {year} {2020})}\BibitemShut {NoStop}%
\bibitem [{\citenamefont {Cui}\ \emph {et~al.}(2017)\citenamefont {Cui},
  \citenamefont {Zheng}, \citenamefont {Ran}, \citenamefont {Wen},
  \citenamefont {Liu}, \citenamefont {Liu}, \citenamefont {Guo},\ and\
  \citenamefont {Yu}}]{Cui2017}%
  \BibitemOpen
  \bibfield  {author} {\bibinfo {author} {\bibfnamefont {Y.}~\bibnamefont
  {Cui}}, \bibinfo {author} {\bibfnamefont {J.}~\bibnamefont {Zheng}}, \bibinfo
  {author} {\bibfnamefont {K.}~\bibnamefont {Ran}}, \bibinfo {author}
  {\bibfnamefont {J.}~\bibnamefont {Wen}}, \bibinfo {author} {\bibfnamefont
  {Z.-X.}\ \bibnamefont {Liu}}, \bibinfo {author} {\bibfnamefont
  {B.}~\bibnamefont {Liu}}, \bibinfo {author} {\bibfnamefont {W.}~\bibnamefont
  {Guo}},\ and\ \bibinfo {author} {\bibfnamefont {W.}~\bibnamefont {Yu}},\
  }\href {https://doi.org/10.1103/PhysRevB.96.205147} {\bibfield  {journal}
  {\bibinfo  {journal} {Phys. Rev. B}\ }\textbf {\bibinfo {volume} {96}},\
  \bibinfo {pages} {205147} (\bibinfo {year} {2017})}\BibitemShut {NoStop}%
\bibitem [{\citenamefont {He}\ \emph {et~al.}(2018)\citenamefont {He},
  \citenamefont {Wang}, \citenamefont {Wang}, \citenamefont {Hardy},
  \citenamefont {Wolf}, \citenamefont {Adelmann}, \citenamefont {Brückel},
  \citenamefont {Su},\ and\ \citenamefont {Meingast}}]{He2018}%
  \BibitemOpen
  \bibfield  {author} {\bibinfo {author} {\bibfnamefont {M.}~\bibnamefont
  {He}}, \bibinfo {author} {\bibfnamefont {X.}~\bibnamefont {Wang}}, \bibinfo
  {author} {\bibfnamefont {L.}~\bibnamefont {Wang}}, \bibinfo {author}
  {\bibfnamefont {F.}~\bibnamefont {Hardy}}, \bibinfo {author} {\bibfnamefont
  {T.}~\bibnamefont {Wolf}}, \bibinfo {author} {\bibfnamefont {P.}~\bibnamefont
  {Adelmann}}, \bibinfo {author} {\bibfnamefont {T.}~\bibnamefont {Brückel}},
  \bibinfo {author} {\bibfnamefont {Y.}~\bibnamefont {Su}},\ and\ \bibinfo
  {author} {\bibfnamefont {C.}~\bibnamefont {Meingast}},\ }\href
  {https://doi.org/10.1088/1361-648X/aada1e} {\bibfield  {journal} {\bibinfo
  {journal} {J.Phys. Conden. Matter}\ }\textbf {\bibinfo {volume} {30}},\
  \bibinfo {pages} {385702} (\bibinfo {year} {2018})}\BibitemShut {NoStop}%
\bibitem [{\citenamefont {Wang}\ \emph {et~al.}(2018)\citenamefont {Wang},
  \citenamefont {Guo}, \citenamefont {Tafti}, \citenamefont {Hegg},
  \citenamefont {Sen}, \citenamefont {Sidorov}, \citenamefont {Wang},
  \citenamefont {Cai}, \citenamefont {Yi}, \citenamefont {Zhou}, \citenamefont
  {Wang}, \citenamefont {Zhang}, \citenamefont {Yang}, \citenamefont {Li},
  \citenamefont {Li}, \citenamefont {Li}, \citenamefont {Liu}, \citenamefont
  {Shi}, \citenamefont {Ku}, \citenamefont {Wu}, \citenamefont {Cava},\ and\
  \citenamefont {Sun}}]{Wang2018}%
  \BibitemOpen
  \bibfield  {author} {\bibinfo {author} {\bibfnamefont {Z.}~\bibnamefont
  {Wang}}, \bibinfo {author} {\bibfnamefont {J.}~\bibnamefont {Guo}}, \bibinfo
  {author} {\bibfnamefont {F.~F.}\ \bibnamefont {Tafti}}, \bibinfo {author}
  {\bibfnamefont {A.}~\bibnamefont {Hegg}}, \bibinfo {author} {\bibfnamefont
  {S.}~\bibnamefont {Sen}}, \bibinfo {author} {\bibfnamefont {V.~A.}\
  \bibnamefont {Sidorov}}, \bibinfo {author} {\bibfnamefont {L.}~\bibnamefont
  {Wang}}, \bibinfo {author} {\bibfnamefont {S.}~\bibnamefont {Cai}}, \bibinfo
  {author} {\bibfnamefont {W.}~\bibnamefont {Yi}}, \bibinfo {author}
  {\bibfnamefont {Y.}~\bibnamefont {Zhou}}, \bibinfo {author} {\bibfnamefont
  {H.}~\bibnamefont {Wang}}, \bibinfo {author} {\bibfnamefont {S.}~\bibnamefont
  {Zhang}}, \bibinfo {author} {\bibfnamefont {K.}~\bibnamefont {Yang}},
  \bibinfo {author} {\bibfnamefont {A.}~\bibnamefont {Li}}, \bibinfo {author}
  {\bibfnamefont {X.}~\bibnamefont {Li}}, \bibinfo {author} {\bibfnamefont
  {Y.}~\bibnamefont {Li}}, \bibinfo {author} {\bibfnamefont {J.}~\bibnamefont
  {Liu}}, \bibinfo {author} {\bibfnamefont {Y.}~\bibnamefont {Shi}}, \bibinfo
  {author} {\bibfnamefont {W.}~\bibnamefont {Ku}}, \bibinfo {author}
  {\bibfnamefont {Q.}~\bibnamefont {Wu}}, \bibinfo {author} {\bibfnamefont
  {R.~J.}\ \bibnamefont {Cava}},\ and\ \bibinfo {author} {\bibfnamefont
  {L.}~\bibnamefont {Sun}},\ }\href
  {https://doi.org/10.1103/PhysRevB.97.245149} {\bibfield  {journal} {\bibinfo
  {journal} {Phys. Rev. B}\ }\textbf {\bibinfo {volume} {97}},\ \bibinfo
  {pages} {245149} (\bibinfo {year} {2018})}\BibitemShut {NoStop}%
\bibitem [{\citenamefont {Bastien}\ \emph {et~al.}(2018)\citenamefont
  {Bastien}, \citenamefont {Garbarino}, \citenamefont {Yadav}, \citenamefont
  {Martinez-Casado}, \citenamefont {Beltr\'an~Rodr\'{\i}guez}, \citenamefont
  {Stahl}, \citenamefont {Kusch}, \citenamefont {Limandri}, \citenamefont
  {Ray}, \citenamefont {Lampen-Kelley}, \citenamefont {Mandrus}, \citenamefont
  {Nagler}, \citenamefont {Roslova}, \citenamefont {Isaeva}, \citenamefont
  {Doert}, \citenamefont {Hozoi}, \citenamefont {Wolter}, \citenamefont
  {B\"uchner}, \citenamefont {Geck},\ and\ \citenamefont {van~den
  Brink}}]{Bastien2018}%
  \BibitemOpen
  \bibfield  {author} {\bibinfo {author} {\bibfnamefont {G.}~\bibnamefont
  {Bastien}}, \bibinfo {author} {\bibfnamefont {G.}~\bibnamefont {Garbarino}},
  \bibinfo {author} {\bibfnamefont {R.}~\bibnamefont {Yadav}}, \bibinfo
  {author} {\bibfnamefont {F.~J.}\ \bibnamefont {Martinez-Casado}}, \bibinfo
  {author} {\bibfnamefont {R.}~\bibnamefont {Beltr\'an~Rodr\'{\i}guez}},
  \bibinfo {author} {\bibfnamefont {Q.}~\bibnamefont {Stahl}}, \bibinfo
  {author} {\bibfnamefont {M.}~\bibnamefont {Kusch}}, \bibinfo {author}
  {\bibfnamefont {S.~P.}\ \bibnamefont {Limandri}}, \bibinfo {author}
  {\bibfnamefont {R.}~\bibnamefont {Ray}}, \bibinfo {author} {\bibfnamefont
  {P.}~\bibnamefont {Lampen-Kelley}}, \bibinfo {author} {\bibfnamefont {D.~G.}\
  \bibnamefont {Mandrus}}, \bibinfo {author} {\bibfnamefont {S.~E.}\
  \bibnamefont {Nagler}}, \bibinfo {author} {\bibfnamefont {M.}~\bibnamefont
  {Roslova}}, \bibinfo {author} {\bibfnamefont {A.}~\bibnamefont {Isaeva}},
  \bibinfo {author} {\bibfnamefont {T.}~\bibnamefont {Doert}}, \bibinfo
  {author} {\bibfnamefont {L.}~\bibnamefont {Hozoi}}, \bibinfo {author}
  {\bibfnamefont {A.~U.~B.}\ \bibnamefont {Wolter}}, \bibinfo {author}
  {\bibfnamefont {B.}~\bibnamefont {B\"uchner}}, \bibinfo {author}
  {\bibfnamefont {J.}~\bibnamefont {Geck}},\ and\ \bibinfo {author}
  {\bibfnamefont {J.}~\bibnamefont {van~den Brink}},\ }\href
  {https://doi.org/10.1103/PhysRevB.97.241108} {\bibfield  {journal} {\bibinfo
  {journal} {Phys. Rev. B}\ }\textbf {\bibinfo {volume} {97}},\ \bibinfo
  {pages} {241108} (\bibinfo {year} {2018})}\BibitemShut {NoStop}%
\bibitem [{\citenamefont {Biesner}\ \emph {et~al.}(2018)\citenamefont
  {Biesner}, \citenamefont {Biswas}, \citenamefont {Li}, \citenamefont {Saito},
  \citenamefont {Pustogow}, \citenamefont {Altmeyer}, \citenamefont {Wolter},
  \citenamefont {B\"uchner}, \citenamefont {Roslova}, \citenamefont {Doert},
  \citenamefont {Winter}, \citenamefont {Valent\'{\i}},\ and\ \citenamefont
  {Dressel}}]{Biesner2018}%
  \BibitemOpen
  \bibfield  {author} {\bibinfo {author} {\bibfnamefont {T.}~\bibnamefont
  {Biesner}}, \bibinfo {author} {\bibfnamefont {S.}~\bibnamefont {Biswas}},
  \bibinfo {author} {\bibfnamefont {W.}~\bibnamefont {Li}}, \bibinfo {author}
  {\bibfnamefont {Y.}~\bibnamefont {Saito}}, \bibinfo {author} {\bibfnamefont
  {A.}~\bibnamefont {Pustogow}}, \bibinfo {author} {\bibfnamefont
  {M.}~\bibnamefont {Altmeyer}}, \bibinfo {author} {\bibfnamefont {A.~U.~B.}\
  \bibnamefont {Wolter}}, \bibinfo {author} {\bibfnamefont {B.}~\bibnamefont
  {B\"uchner}}, \bibinfo {author} {\bibfnamefont {M.}~\bibnamefont {Roslova}},
  \bibinfo {author} {\bibfnamefont {T.}~\bibnamefont {Doert}}, \bibinfo
  {author} {\bibfnamefont {S.~M.}\ \bibnamefont {Winter}}, \bibinfo {author}
  {\bibfnamefont {R.}~\bibnamefont {Valent\'{\i}}},\ and\ \bibinfo {author}
  {\bibfnamefont {M.}~\bibnamefont {Dressel}},\ }\href
  {https://doi.org/10.1103/PhysRevB.97.220401} {\bibfield  {journal} {\bibinfo
  {journal} {Phys. Rev. B}\ }\textbf {\bibinfo {volume} {97}},\ \bibinfo
  {pages} {220401} (\bibinfo {year} {2018})}\BibitemShut {NoStop}%
\bibitem [{\citenamefont {Li}\ \emph {et~al.}(2019)\citenamefont {Li},
  \citenamefont {Chen}, \citenamefont {Gan}, \citenamefont {Li}, \citenamefont
  {Yan}, \citenamefont {Ye}, \citenamefont {Pei}, \citenamefont {Zhang},
  \citenamefont {Wang}, \citenamefont {Su}, \citenamefont {Dai}, \citenamefont
  {Chen}, \citenamefont {Shi}, \citenamefont {Wang}, \citenamefont {Zhang},
  \citenamefont {Wang}, \citenamefont {Yu}, \citenamefont {Ye}, \citenamefont
  {Mei},\ and\ \citenamefont {Huang}}]{Li2019}%
  \BibitemOpen
  \bibfield  {author} {\bibinfo {author} {\bibfnamefont {G.}~\bibnamefont
  {Li}}, \bibinfo {author} {\bibfnamefont {X.}~\bibnamefont {Chen}}, \bibinfo
  {author} {\bibfnamefont {Y.}~\bibnamefont {Gan}}, \bibinfo {author}
  {\bibfnamefont {F.}~\bibnamefont {Li}}, \bibinfo {author} {\bibfnamefont
  {M.}~\bibnamefont {Yan}}, \bibinfo {author} {\bibfnamefont {F.}~\bibnamefont
  {Ye}}, \bibinfo {author} {\bibfnamefont {S.}~\bibnamefont {Pei}}, \bibinfo
  {author} {\bibfnamefont {Y.}~\bibnamefont {Zhang}}, \bibinfo {author}
  {\bibfnamefont {L.}~\bibnamefont {Wang}}, \bibinfo {author} {\bibfnamefont
  {H.}~\bibnamefont {Su}}, \bibinfo {author} {\bibfnamefont {J.}~\bibnamefont
  {Dai}}, \bibinfo {author} {\bibfnamefont {Y.}~\bibnamefont {Chen}}, \bibinfo
  {author} {\bibfnamefont {Y.}~\bibnamefont {Shi}}, \bibinfo {author}
  {\bibfnamefont {X.}~\bibnamefont {Wang}}, \bibinfo {author} {\bibfnamefont
  {L.}~\bibnamefont {Zhang}}, \bibinfo {author} {\bibfnamefont
  {S.}~\bibnamefont {Wang}}, \bibinfo {author} {\bibfnamefont {D.}~\bibnamefont
  {Yu}}, \bibinfo {author} {\bibfnamefont {F.}~\bibnamefont {Ye}}, \bibinfo
  {author} {\bibfnamefont {J.-W.}\ \bibnamefont {Mei}},\ and\ \bibinfo {author}
  {\bibfnamefont {M.}~\bibnamefont {Huang}},\ }\href
  {https://doi.org/10.1103/PhysRevMaterials.3.023601} {\bibfield  {journal}
  {\bibinfo  {journal} {Phys. Rev. Materials}\ }\textbf {\bibinfo {volume}
  {3}},\ \bibinfo {pages} {023601} (\bibinfo {year} {2019})}\BibitemShut
  {NoStop}%
\bibitem [{\citenamefont {Wolf}\ \emph {et~al.}(2022)\citenamefont {Wolf},
  \citenamefont {Kaib}, \citenamefont {Razpopov}, \citenamefont {Biswas},
  \citenamefont {Riedl}, \citenamefont {Winter}, \citenamefont {Valent\'{\i}},
  \citenamefont {Saito}, \citenamefont {Hartmann}, \citenamefont {Vinokurova},
  \citenamefont {Doert}, \citenamefont {Isaeva}, \citenamefont {Bastien},
  \citenamefont {Wolter}, \citenamefont {B\"uchner},\ and\ \citenamefont
  {Lang}}]{Wolf2022}%
  \BibitemOpen
  \bibfield  {author} {\bibinfo {author} {\bibfnamefont {B.}~\bibnamefont
  {Wolf}}, \bibinfo {author} {\bibfnamefont {D.~A.~S.}\ \bibnamefont {Kaib}},
  \bibinfo {author} {\bibfnamefont {A.}~\bibnamefont {Razpopov}}, \bibinfo
  {author} {\bibfnamefont {S.}~\bibnamefont {Biswas}}, \bibinfo {author}
  {\bibfnamefont {K.}~\bibnamefont {Riedl}}, \bibinfo {author} {\bibfnamefont
  {S.~M.}\ \bibnamefont {Winter}}, \bibinfo {author} {\bibfnamefont
  {R.}~\bibnamefont {Valent\'{\i}}}, \bibinfo {author} {\bibfnamefont
  {Y.}~\bibnamefont {Saito}}, \bibinfo {author} {\bibfnamefont
  {S.}~\bibnamefont {Hartmann}}, \bibinfo {author} {\bibfnamefont
  {E.}~\bibnamefont {Vinokurova}}, \bibinfo {author} {\bibfnamefont
  {T.}~\bibnamefont {Doert}}, \bibinfo {author} {\bibfnamefont
  {A.}~\bibnamefont {Isaeva}}, \bibinfo {author} {\bibfnamefont
  {G.}~\bibnamefont {Bastien}}, \bibinfo {author} {\bibfnamefont {A.~U.~B.}\
  \bibnamefont {Wolter}}, \bibinfo {author} {\bibfnamefont {B.}~\bibnamefont
  {B\"uchner}},\ and\ \bibinfo {author} {\bibfnamefont {M.}~\bibnamefont
  {Lang}},\ }\href {https://doi.org/10.1103/PhysRevB.106.134432} {\bibfield
  {journal} {\bibinfo  {journal} {Phys. Rev. B}\ }\textbf {\bibinfo {volume}
  {106}},\ \bibinfo {pages} {134432} (\bibinfo {year} {2022})}\BibitemShut
  {NoStop}%
\bibitem [{\citenamefont {Mi}\ \emph {et~al.}(2021)\citenamefont {Mi},
  \citenamefont {Wang}, \citenamefont {Gui}, \citenamefont {Pi}, \citenamefont
  {Zheng}, \citenamefont {Yang}, \citenamefont {Gan}, \citenamefont {Wang},
  \citenamefont {Li}, \citenamefont {Wang}, \citenamefont {Zhang},
  \citenamefont {Su}, \citenamefont {Chai},\ and\ \citenamefont {He}}]{Mi2021}%
  \BibitemOpen
  \bibfield  {author} {\bibinfo {author} {\bibfnamefont {X.}~\bibnamefont
  {Mi}}, \bibinfo {author} {\bibfnamefont {X.}~\bibnamefont {Wang}}, \bibinfo
  {author} {\bibfnamefont {H.}~\bibnamefont {Gui}}, \bibinfo {author}
  {\bibfnamefont {M.}~\bibnamefont {Pi}}, \bibinfo {author} {\bibfnamefont
  {T.}~\bibnamefont {Zheng}}, \bibinfo {author} {\bibfnamefont
  {K.}~\bibnamefont {Yang}}, \bibinfo {author} {\bibfnamefont {Y.}~\bibnamefont
  {Gan}}, \bibinfo {author} {\bibfnamefont {P.}~\bibnamefont {Wang}}, \bibinfo
  {author} {\bibfnamefont {A.}~\bibnamefont {Li}}, \bibinfo {author}
  {\bibfnamefont {A.}~\bibnamefont {Wang}}, \bibinfo {author} {\bibfnamefont
  {L.}~\bibnamefont {Zhang}}, \bibinfo {author} {\bibfnamefont
  {Y.}~\bibnamefont {Su}}, \bibinfo {author} {\bibfnamefont {Y.}~\bibnamefont
  {Chai}},\ and\ \bibinfo {author} {\bibfnamefont {M.}~\bibnamefont {He}},\
  }\href {https://doi.org/10.1103/PhysRevB.103.174413} {\bibfield  {journal}
  {\bibinfo  {journal} {Phys. Rev. B}\ }\textbf {\bibinfo {volume} {103}},\
  \bibinfo {pages} {174413} (\bibinfo {year} {2021})}\BibitemShut {NoStop}%
\bibitem [{\citenamefont {Park}\ \emph {et~al.}(2016)\citenamefont {Park},
  \citenamefont {Do}, \citenamefont {Choi}, \citenamefont {Jang}, \citenamefont
  {Jang}, \citenamefont {Schefer}, \citenamefont {Wu}, \citenamefont {Gardner},
  \citenamefont {Park}, \citenamefont {Park},\ and\ \citenamefont
  {Ji}}]{Park2016}%
  \BibitemOpen
  \bibfield  {author} {\bibinfo {author} {\bibfnamefont {S.~Y.}\ \bibnamefont
  {Park}}, \bibinfo {author} {\bibfnamefont {S.~H.}\ \bibnamefont {Do}},
  \bibinfo {author} {\bibfnamefont {K.~Y.}\ \bibnamefont {Choi}}, \bibinfo
  {author} {\bibfnamefont {D.}~\bibnamefont {Jang}}, \bibinfo {author}
  {\bibfnamefont {T.~H.}\ \bibnamefont {Jang}}, \bibinfo {author}
  {\bibfnamefont {J.}~\bibnamefont {Schefer}}, \bibinfo {author} {\bibfnamefont
  {C.~M.}\ \bibnamefont {Wu}}, \bibinfo {author} {\bibfnamefont {J.~S.}\
  \bibnamefont {Gardner}}, \bibinfo {author} {\bibfnamefont {J.~M.~S.}\
  \bibnamefont {Park}}, \bibinfo {author} {\bibfnamefont {J.~H.}\ \bibnamefont
  {Park}},\ and\ \bibinfo {author} {\bibfnamefont {S.}~\bibnamefont {Ji}},\
  }\href@noop {} {\bibfield  {journal} {\bibinfo  {journal} {arXiv:1609.05690}\
  } (\bibinfo {year} {2016})}\BibitemShut {NoStop}%
\bibitem [{\citenamefont {Johnson}\ \emph {et~al.}(2015)\citenamefont
  {Johnson}, \citenamefont {Williams}, \citenamefont {Haghighirad},
  \citenamefont {Singleton}, \citenamefont {Zapf}, \citenamefont {Manuel},
  \citenamefont {Mazin}, \citenamefont {Li}, \citenamefont {Jeschke},
  \citenamefont {Valent\'{\i}},\ and\ \citenamefont {Coldea}}]{Johnson2015}%
  \BibitemOpen
  \bibfield  {author} {\bibinfo {author} {\bibfnamefont {R.~D.}\ \bibnamefont
  {Johnson}}, \bibinfo {author} {\bibfnamefont {S.~C.}\ \bibnamefont
  {Williams}}, \bibinfo {author} {\bibfnamefont {A.~A.}\ \bibnamefont
  {Haghighirad}}, \bibinfo {author} {\bibfnamefont {J.}~\bibnamefont
  {Singleton}}, \bibinfo {author} {\bibfnamefont {V.}~\bibnamefont {Zapf}},
  \bibinfo {author} {\bibfnamefont {P.}~\bibnamefont {Manuel}}, \bibinfo
  {author} {\bibfnamefont {I.~I.}\ \bibnamefont {Mazin}}, \bibinfo {author}
  {\bibfnamefont {Y.}~\bibnamefont {Li}}, \bibinfo {author} {\bibfnamefont
  {H.~O.}\ \bibnamefont {Jeschke}}, \bibinfo {author} {\bibfnamefont
  {R.}~\bibnamefont {Valent\'{\i}}},\ and\ \bibinfo {author} {\bibfnamefont
  {R.}~\bibnamefont {Coldea}},\ }\href
  {https://doi.org/10.1103/PhysRevB.92.235119} {\bibfield  {journal} {\bibinfo
  {journal} {Phys. Rev. B}\ }\textbf {\bibinfo {volume} {92}},\ \bibinfo
  {pages} {235119} (\bibinfo {year} {2015})}\BibitemShut {NoStop}%
\bibitem [{\citenamefont {Cao}\ \emph {et~al.}(2016)\citenamefont {Cao},
  \citenamefont {Banerjee}, \citenamefont {Yan}, \citenamefont {Bridges},
  \citenamefont {Lumsden}, \citenamefont {Mandrus}, \citenamefont {Tennant},
  \citenamefont {Chakoumakos},\ and\ \citenamefont {Nagler}}]{Cao2016}%
  \BibitemOpen
  \bibfield  {author} {\bibinfo {author} {\bibfnamefont {H.~B.}\ \bibnamefont
  {Cao}}, \bibinfo {author} {\bibfnamefont {A.}~\bibnamefont {Banerjee}},
  \bibinfo {author} {\bibfnamefont {J.-Q.}\ \bibnamefont {Yan}}, \bibinfo
  {author} {\bibfnamefont {C.~A.}\ \bibnamefont {Bridges}}, \bibinfo {author}
  {\bibfnamefont {M.~D.}\ \bibnamefont {Lumsden}}, \bibinfo {author}
  {\bibfnamefont {D.~G.}\ \bibnamefont {Mandrus}}, \bibinfo {author}
  {\bibfnamefont {D.~A.}\ \bibnamefont {Tennant}}, \bibinfo {author}
  {\bibfnamefont {B.~C.}\ \bibnamefont {Chakoumakos}},\ and\ \bibinfo {author}
  {\bibfnamefont {S.~E.}\ \bibnamefont {Nagler}},\ }\href
  {https://doi.org/10.1103/PhysRevB.93.134423} {\bibfield  {journal} {\bibinfo
  {journal} {Phys. Rev. B}\ }\textbf {\bibinfo {volume} {93}},\ \bibinfo
  {pages} {134423} (\bibinfo {year} {2016})}\BibitemShut {NoStop}%
\bibitem [{\citenamefont {Kubota}\ \emph {et~al.}(2015)\citenamefont {Kubota},
  \citenamefont {Tanaka}, \citenamefont {Ono}, \citenamefont {Narumi},\ and\
  \citenamefont {Kindo}}]{Kubota2015}%
  \BibitemOpen
  \bibfield  {author} {\bibinfo {author} {\bibfnamefont {Y.}~\bibnamefont
  {Kubota}}, \bibinfo {author} {\bibfnamefont {H.}~\bibnamefont {Tanaka}},
  \bibinfo {author} {\bibfnamefont {T.}~\bibnamefont {Ono}}, \bibinfo {author}
  {\bibfnamefont {Y.}~\bibnamefont {Narumi}},\ and\ \bibinfo {author}
  {\bibfnamefont {K.}~\bibnamefont {Kindo}},\ }\href
  {https://doi.org/10.1103/PhysRevB.91.094422} {\bibfield  {journal} {\bibinfo
  {journal} {Phys. Rev. B}\ }\textbf {\bibinfo {volume} {91}},\ \bibinfo
  {pages} {094422} (\bibinfo {year} {2015})}\BibitemShut {NoStop}%
\bibitem [{\citenamefont {Zhou}\ \emph {et~al.}(2019)\citenamefont {Zhou},
  \citenamefont {Wang}, \citenamefont {Osterhoudt}, \citenamefont
  {Lampen-Kelley}, \citenamefont {Mandrus}, \citenamefont {He}, \citenamefont
  {Burch},\ and\ \citenamefont {Henriksen}}]{Zhou2019}%
  \BibitemOpen
  \bibfield  {author} {\bibinfo {author} {\bibfnamefont {B.}~\bibnamefont
  {Zhou}}, \bibinfo {author} {\bibfnamefont {Y.}~\bibnamefont {Wang}}, \bibinfo
  {author} {\bibfnamefont {G.~B.}\ \bibnamefont {Osterhoudt}}, \bibinfo
  {author} {\bibfnamefont {P.}~\bibnamefont {Lampen-Kelley}}, \bibinfo {author}
  {\bibfnamefont {D.}~\bibnamefont {Mandrus}}, \bibinfo {author} {\bibfnamefont
  {R.}~\bibnamefont {He}}, \bibinfo {author} {\bibfnamefont {K.~S.}\
  \bibnamefont {Burch}},\ and\ \bibinfo {author} {\bibfnamefont {E.~A.}\
  \bibnamefont {Henriksen}},\ }\href
  {https://doi.org/https://doi.org/10.1016/j.jpcs.2018.01.026} {\bibfield
  {journal} {\bibinfo  {journal} {J. Phys. Chem. Solids}\ }\textbf {\bibinfo
  {volume} {128}},\ \bibinfo {pages} {291} (\bibinfo {year}
  {2019})}\BibitemShut {NoStop}%
\bibitem [{\citenamefont {Gass}\ \emph {et~al.}(2020)\citenamefont {Gass},
  \citenamefont {C\^onsoli}, \citenamefont {Kocsis}, \citenamefont {Corredor},
  \citenamefont {Lampen-Kelley}, \citenamefont {Mandrus}, \citenamefont
  {Nagler}, \citenamefont {Janssen}, \citenamefont {Vojta}, \citenamefont
  {B\"uchner},\ and\ \citenamefont {Wolter}}]{Gass2022}%
  \BibitemOpen
  \bibfield  {author} {\bibinfo {author} {\bibfnamefont {S.}~\bibnamefont
  {Gass}}, \bibinfo {author} {\bibfnamefont {P.~M.}\ \bibnamefont {C\^onsoli}},
  \bibinfo {author} {\bibfnamefont {V.}~\bibnamefont {Kocsis}}, \bibinfo
  {author} {\bibfnamefont {L.~T.}\ \bibnamefont {Corredor}}, \bibinfo {author}
  {\bibfnamefont {P.}~\bibnamefont {Lampen-Kelley}}, \bibinfo {author}
  {\bibfnamefont {D.~G.}\ \bibnamefont {Mandrus}}, \bibinfo {author}
  {\bibfnamefont {S.~E.}\ \bibnamefont {Nagler}}, \bibinfo {author}
  {\bibfnamefont {L.}~\bibnamefont {Janssen}}, \bibinfo {author} {\bibfnamefont
  {M.}~\bibnamefont {Vojta}}, \bibinfo {author} {\bibfnamefont
  {B.}~\bibnamefont {B\"uchner}},\ and\ \bibinfo {author} {\bibfnamefont
  {A.~U.~B.}\ \bibnamefont {Wolter}},\ }\href
  {https://doi.org/10.1103/PhysRevB.101.245158} {\bibfield  {journal} {\bibinfo
   {journal} {Phys. Rev. B}\ }\textbf {\bibinfo {volume} {101}},\ \bibinfo
  {pages} {245158} (\bibinfo {year} {2020})}\BibitemShut {NoStop}%
\bibitem [{\citenamefont {Reschke}\ \emph {et~al.}(2018)\citenamefont
  {Reschke}, \citenamefont {Mayr}, \citenamefont {Widmann}, \citenamefont {von
  Nidda}, \citenamefont {Tsurkan}, \citenamefont {Eremin}, \citenamefont {Do},
  \citenamefont {Choi}, \citenamefont {Wang},\ and\ \citenamefont
  {Loidl}}]{Reschke2018}%
  \BibitemOpen
  \bibfield  {author} {\bibinfo {author} {\bibfnamefont {S.}~\bibnamefont
  {Reschke}}, \bibinfo {author} {\bibfnamefont {F.}~\bibnamefont {Mayr}},
  \bibinfo {author} {\bibfnamefont {S.}~\bibnamefont {Widmann}}, \bibinfo
  {author} {\bibfnamefont {H.-A.~K.}\ \bibnamefont {von Nidda}}, \bibinfo
  {author} {\bibfnamefont {V.}~\bibnamefont {Tsurkan}}, \bibinfo {author}
  {\bibfnamefont {M.~V.}\ \bibnamefont {Eremin}}, \bibinfo {author}
  {\bibfnamefont {S.-H.}\ \bibnamefont {Do}}, \bibinfo {author} {\bibfnamefont
  {K.-Y.}\ \bibnamefont {Choi}}, \bibinfo {author} {\bibfnamefont
  {Z.}~\bibnamefont {Wang}},\ and\ \bibinfo {author} {\bibfnamefont
  {A.}~\bibnamefont {Loidl}},\ }\href
  {https://doi.org/10.1088/1361-648X/aae805} {\bibfield  {journal} {\bibinfo
  {journal} {Journal of Physics: Condensed Matter}\ }\textbf {\bibinfo {volume}
  {30}},\ \bibinfo {pages} {475604} (\bibinfo {year} {2018})}\BibitemShut
  {NoStop}%
\bibitem [{\citenamefont {Shen}\ \emph {et~al.}(2022)\citenamefont {Shen},
  \citenamefont {Breitner}, \citenamefont {Prishchenko}, \citenamefont {Manna},
  \citenamefont {Jesche}, \citenamefont {Seidler}, \citenamefont {Gegenwart},\
  and\ \citenamefont {Tsirlin}}]{Shen2022}%
  \BibitemOpen
  \bibfield  {author} {\bibinfo {author} {\bibfnamefont {B.}~\bibnamefont
  {Shen}}, \bibinfo {author} {\bibfnamefont {F.}~\bibnamefont {Breitner}},
  \bibinfo {author} {\bibfnamefont {D.}~\bibnamefont {Prishchenko}}, \bibinfo
  {author} {\bibfnamefont {R.~S.}\ \bibnamefont {Manna}}, \bibinfo {author}
  {\bibfnamefont {A.}~\bibnamefont {Jesche}}, \bibinfo {author} {\bibfnamefont
  {M.~L.}\ \bibnamefont {Seidler}}, \bibinfo {author} {\bibfnamefont
  {P.}~\bibnamefont {Gegenwart}},\ and\ \bibinfo {author} {\bibfnamefont
  {A.~A.}\ \bibnamefont {Tsirlin}},\ }\href
  {https://doi.org/10.1103/PhysRevB.105.054412} {\bibfield  {journal} {\bibinfo
   {journal} {Phys. Rev. B}\ }\textbf {\bibinfo {volume} {105}},\ \bibinfo
  {pages} {054412} (\bibinfo {year} {2022})}\BibitemShut {NoStop}%
\bibitem [{\citenamefont {Takayama}\ \emph {et~al.}(2019)\citenamefont
  {Takayama}, \citenamefont {Krajewska}, \citenamefont {Gibbs}, \citenamefont
  {Yaresko}, \citenamefont {Ishii}, \citenamefont {Yamaoka}, \citenamefont
  {Ishii}, \citenamefont {Hiraoka}, \citenamefont {Funnell}, \citenamefont
  {Bull},\ and\ \citenamefont {Takagi}}]{Takayama2019}%
  \BibitemOpen
  \bibfield  {author} {\bibinfo {author} {\bibfnamefont {T.}~\bibnamefont
  {Takayama}}, \bibinfo {author} {\bibfnamefont {A.}~\bibnamefont {Krajewska}},
  \bibinfo {author} {\bibfnamefont {A.~S.}\ \bibnamefont {Gibbs}}, \bibinfo
  {author} {\bibfnamefont {A.~N.}\ \bibnamefont {Yaresko}}, \bibinfo {author}
  {\bibfnamefont {H.}~\bibnamefont {Ishii}}, \bibinfo {author} {\bibfnamefont
  {H.}~\bibnamefont {Yamaoka}}, \bibinfo {author} {\bibfnamefont
  {K.}~\bibnamefont {Ishii}}, \bibinfo {author} {\bibfnamefont
  {N.}~\bibnamefont {Hiraoka}}, \bibinfo {author} {\bibfnamefont {N.~P.}\
  \bibnamefont {Funnell}}, \bibinfo {author} {\bibfnamefont {C.~L.}\
  \bibnamefont {Bull}},\ and\ \bibinfo {author} {\bibfnamefont
  {H.}~\bibnamefont {Takagi}},\ }\href
  {https://doi.org/10.1103/PhysRevB.99.125127} {\bibfield  {journal} {\bibinfo
  {journal} {Phys. Rev. B}\ }\textbf {\bibinfo {volume} {99}},\ \bibinfo
  {pages} {125127} (\bibinfo {year} {2019})}\BibitemShut {NoStop}%
\bibitem [{\citenamefont {Liu}\ \emph {et~al.}(2022)\citenamefont {Liu},
  \citenamefont {Chaloupka},\ and\ \citenamefont {Khaliullin}}]{Liu2022}%
  \BibitemOpen
  \bibfield  {author} {\bibinfo {author} {\bibfnamefont {H.}~\bibnamefont
  {Liu}}, \bibinfo {author} {\bibfnamefont {J.}~\bibnamefont {Chaloupka}},\
  and\ \bibinfo {author} {\bibfnamefont {G.}~\bibnamefont {Khaliullin}},\
  }\href {https://doi.org/10.1103/PhysRevB.105.214411} {\bibfield  {journal}
  {\bibinfo  {journal} {Phys. Rev. B}\ }\textbf {\bibinfo {volume} {105}},\
  \bibinfo {pages} {214411} (\bibinfo {year} {2022})}\BibitemShut {NoStop}%
\end{thebibliography}%

\end{document}